\documentclass[11pt, a4paper]{article}
\usepackage{jstyle, enumitem}
\usepackage{hyperref, subcaption}
\usepackage{slashed,framed}
\usepackage{empheq}
\usetikzlibrary{calc,decorations.markings}
\usepackage[normalem]{ulem}

\usepackage{amsfonts, amssymb, amsmath, amsthm, stmaryrd, amsbsy,
  mathrsfs, genyoungtabtikz, mathdots, dsfont, relsize}
\usepackage{color, fancybox, empheq, graphicx}
\usepackage{mdframed}
\usepackage[strict]{changepage}

\newcommand{\so}{\mathfrak{so}}

\renewcommand{\D}{\mathcal{D}}
\newcommand{\N}{\mathbb{N}}
\newcommand{\R}{\mathbb{R}}
\renewcommand{\cZ}{\mathcal{Z}}
\newcommand{\Y}{\mathbb{Y}}
\newcommand{\V}{\mathcal{V}}
\newcommand{\Vp}{\mathbb{V}}
\renewcommand{\U}{\mathcal{U}}
\newcommand{\Sh}{\mathcal{S}}
\newcommand{\dd}{\mathrm{d}}
\newcommand{\rac}{\mathrm{Rac}}
\newcommand{\di}{\mathrm{Di}}

\newcommand{\zero}{\boldsymbol{0}}

\newcommand{\proj}{\mathbb{P}}
\renewcommand{\th}{\text{th}}
\newcommand{\Pd}[1]{\mathcal{P}_{#1}}
\newcommand{\pos}[2]{\underset{#2}{\underset{\uparrow}{#1}}}
\newcommand{\1}{\mathbf{1}}
\newcommand{\branch}{\overset{\so(2,d)}{\underset{\so(2,d-1)}{\downarrow}}}
\newcommand{\branchmod}[2]{\overset{\so(2,#1)}{\underset{\so(2,#2)}{\downarrow}}}
\newcommand{\bracket}[1][0pt]{%
  \mathrel{\raisebox{#1}{${\scriptstyle m}\,\bigg\{$}}%
}

\definecolor{theRed}{rgb}{0.56,0,0}

\newtheorem{theorem}{Theorem}[section] \newtheorem*{theorem*}{Theorem}
\newtheorem{lemma}[theorem]{Lemma} \newtheorem*{lemma*}{Lemma}

\theoremstyle{definition}

\newtheorem*{definition*}{Definition}

\Yboxdim{8pt} 
\Ylinethick{0.9pt}

\usepackage{tikz, tikz-cd, animate}
\usetikzlibrary{arrows, matrix}

\author[a]{Thomas BASILE}
\author[b]{\qquad Xavier BEKAERT}
\author[a]{\qquad Euihun JOUNG}

\affiliation[a]{Department of Physics and Research Institute of Basic
  Science, \\
	Kyung Hee University,\\ Seoul 02447, Korea}
\affiliation[b]{Institut Denis Poisson,\\
Universit\'e de Tours, Universit\'e d'Orl\'eans, CNRS,\\
  Parc de Grandmont, 37200 Tours, France}

\emailAdd{thomas.basile@khu.ac.kr}
\emailAdd{xavier.bekaert@lmpt.univ-tours.fr}
\emailAdd{euihun.joung@khu.ac.kr}

\begin{document}

\title{\centering Conformal Higher-Spin Gravity: \\ Linearized
  Spectrum = Symmetry Algebra}

\abstract{  
  The linearized spectrum and the algebra of global symmetries of
  conformal higher-spin gravity decompose into infinitely many
  representations of the conformal algebra. Their characters involve
  divergent sums over spins. We propose a suitable regularization
  adapted to their evaluation and observe that their characters are
  actually equal. This result holds in the case of type-A and type-B
  (and their higher-depth generalizations) theories and confirms
  previous observations on a remarkable rearrangement of dynamical
  degrees of freedom in conformal higher-spin gravity after
  regularization.
}

\maketitle

\section{Introduction}
\label{sec:intro}

Whether Einstein's gravity can be extended to a theory with larger
symmetries than the usual diffeomorphisms is a challenging question
and attempts at answering it have brought many interesting results not
only to the physics of gravitation but also to mathematics. An old
example is provided by the fruitful exchanges between Weyl's theory of
gravity \cite{Weyl:1919fi} and conformal geometry. Higher-spin gravity
is a much younger example and its underlying geometry remains somewhat
elusive. However, its deep connections with important topics of modern
mathematics (such as deformation quantization\footnote{The close
  relationships between higher-spin gravity and deformation
  quantization was observed in various instances over the last three
  decades. It was first observed by Vasiliev in his realisation of
  higher-spin algebras in terms of star products
  \cite{Vasiliev:1986qx}. Aside from this seminal paper, other
  examples are the relation between the quantized hyperboloid and the
  deformed oscillator algebra which is instrumental in Vasiliev
  nonlinear equations \cite{Vasiliev:1989qh, Vasiliev:1989re}, the
  connection of the latter equations with Fedosov quantization
  \cite{Barnich:2004cr, Vasiliev:2005zu, Grigoriev:2006tt} and with
  some formality theorems \cite{Sharapov:2017yde, Sharapov:2017lxr},
  etc.}, jet bundles, tractors, etc) might keep some surprises in
store.

What we shall study in this paper is the theory which combines the
conformal and higher-spin extensions of gravity, namely conformal
higher-spin (CHS) gravity.  The study of CHS gauge fields was started
by Fradkin and Tseytlin in \cite{Fradkin:1985am}, where the
higher-spin generalization of the linearized conformal gravity action
was obtained.  Subsequently, the cubic interactions of CHS fields were
analyzed by Fradkin and Linetsky in a frame-like formulation
\cite{Fradkin:1989md, Fradkin:1990ps} (see also \cite{Fradkin:1989ywa,
  Fradkin:1989yd} for early works on the CHS superalgebra) where the
quartic order analysis nevertheless becomes difficult, similarly to
the Fradkin-Vasiliev construction of the massless higher-spin (MHS)
interactions \cite{Fradkin:1986qy, Fradkin:1987ks}. In three
dimensions however, the full non-linear action could be obtained in
the Chern-Simons formulation \cite{Fradkin:1989xt, Horne:1988jf,
  Pope:1989vj}. The unfolded formulation on which Vasiliev's equations
are based can be also applied to free CHS gauge fields (see
e.g. \cite{Shaynkman:2001ip, Vasiliev:2009ck,
  Nilsson:2013tva,Shaynkman:2014fqa, Nilsson:2015pua, Basile:2017mqc,
  Shaynkman:2018kkc}).

In fact, one of the simplest ways to understand CHS gravity in even
dimensions is viewing it as the logarithmically divergent part of the
effective action of a conformal scalar field in the background of all
higher-spin fields \cite{Tseytlin:2002gz, Segal:2002gd,
  Bekaert:2010ky}, which can be computed perturbatively in the weak
field expansion.  In the context of higher-spin holography, the CHS
action can be again related to (the logarithmically divergent part of)
the on-shell bulk action of the MHS gravity in one higher dimension
with standard boundary conditions. The CHS field equations can also be
obtained as the obstruction in the power series expansion (\`a la
Fefferman-Graham) of the solutions to the bulk MHS field equations
\cite{Bekaert:2012vt, Bekaert:2013zya}.  This holographic picture
makes it clear that the global symmetry of the CHS theory matches that
of MHS theory (see e.g. \cite{Bekaert:2010ky}). This scenario actually
extends beyond the case of the conformal scalar field, the
construction of CHS theories via the effective action ensures that
there is a one-to-one correspondence between free vector model conformal field
theories (CFTs) and CHS theories,\footnote{See
  e.g. \cite{Bonora:2016otz, Bonora:2017ykb}, as well as
  \cite{Bonezzi:2017mwr, Bonora:2018uwx} where the effective actions
  have been calculated in the world-line formulation.} while the
higher-spin holographic duality ensures that there is a one-to-one
correspondence between them and MHS theories. Accordingly, we will
denote the zoo of CHS theories following the by-now standard
nomenclature (type A, B, etc) for MHS theories.\footnote{In the
  sequel, the somewhat ambiguous terms ``CHS theory'' or ``CHS
  gravity'' will stand either for the specific example of type-A CHS
  gravity or, on the contrary, for generic CHS theories. The context
  should make it clear.}  The type-A, -B and -C higher-spin gravities
on AdS$_{d+1}$ are dual to the (singlet sector of) vector model
CFT$_d$ of free scalars, spinors and $\frac{d-2}2$-forms,
respectively.

CHS gravity shows several interesting properties. In particular, it
shares the main drawback and virtues of Weyl's gravity. On the one
hand, it is a higher-derivative theory so it is non-unitary. On the
other hand, it is an interacting theory with a massless spin-two field
in its spectrum whose interactions with the other fields include the
minimal coupling, so it is a theory of gravity. Moreover, its global
symmetries include conformal symmetry thus it is scale-invariant at
the classical level. Consequently, the absence of gauge anomalies (in particular, Weyl anomalies and their higher-spin versions) would
ensure that CHS gravity is UV-finite (see e.g. \cite{Fradkin:1985am}
for the original arguments). In fact, an advantage of CHS gravity over
its spin-two counterpart is that it might be anomaly-free, as
suggested by some preliminary tests.  From a quantum gravity
perspective, this motivates the study of CHS gravity as an interesting
toy model of a UV-finite (albeit non-unitary) theory of quantum
gravity.  The spectra of higher-spin gravity theories (CHS and MHS)
involve infinite towers of gauge fields. The corresponding infinite
number of contributions require some regularization which opens the
possibility that the collective behaviour of such infinite collection
of fields may be much softer than their individual behaviour. In fact,
due to the presence of huge symmetries, such systems have been
observed to exhibit remarkable cancellation properties in their
scattering amplitudes, one-loop anomalies, partition functions,
etc. From a higher-spin theory perspective, CHS gravity can be viewed
as a theory of interacting massless and partially-massless fields of
all depths and all spins. It has the nice feature of being
perturbatively local and of admitting a relatively simple metric-like
formulation.  Moreover, it appears to possess similar features to MHS
theory. In fact, 
non-standard MHS gravity could possibly arise from CHS gravity at tree level
by imposing suitable boundary conditions around
an anti de Sitter (AdS) background, in the same way that Einstein's
gravity is related to Weyl's gravity \cite{Maldacena:2011mk,
  Anastasiou:2016jix}.

The linear equation for the conformal spin-$s$ field is local in even
$d$ dimensions and has $2s+d-4$ derivatives. Generically, the
corresponding kinetic operator is higher-derivative and can be
factorized into the ones for massless and partially-massless fields of
the same spin (around AdS$_d$) as was suggested in \cite{Tseytlin:2013jya,Metsaev:2014iwa,
  Nutma:2014pua,Grigoriev:2018mkp}\,\footnote{The factorization in the
  spin-two case was observed already in \cite{Fradkin:1983zz,
    Tseytlin:1984wj,Deser:1983mm}.} (see also \cite{Joung:2012qy} for related
discussions). This factorization allows us to compute easily its Weyl
anomaly and it was found that the $a$-anomaly coefficient vanishes
\cite{Giombi:2013yva, Tseytlin:2013jya}. The $c$-anomaly would also
vanish assuming a factorized form of the equation in the presence of
non-vanishing curvature \cite{Tseytlin:2013jya}. Unfortunately, such a
factorization does not actually happen in general, as was shown in
\cite{Nutma:2014pua}. As a consequence, to determine whether the
$c$-anomaly vanishes or not, one needs the linear conformal spin-$s$
equation in an arbitrary gravitational background, which has been the
subject of several recent papers \cite{Grigoriev:2016bzl,
  Beccaria:2017nco} (see also \cite{Beccaria:2017lcz, Acevedo:2017vkk}
for other strategies).

Scattering amplitudes of CHS gravity also exhibit surprising features:
four-point scattering amplitudes of external scalar fields, Maxwell
fields and Weyl gravitons mediated by the infinite tower of CHS fields
vanish \cite{Joung:2015eny,Beccaria:2016syk}. There exists yet another
approach to CHS theory using the twistor formalism
\cite{Haehnel:2016mlb}. The twistor CHS theory has the same linear
action as the conventional CHS, but it is not clear that the two CHS
theories are the same at the full interacting level. Recently,
scattering amplitudes of twistor CHS theory have been studied in
\cite{Adamo:2016ple, Adamo:2018srx} (see also \cite{Adamo:2013tja} in
the context of conformal gravity).

In this paper, we will focus on the spectrum and symmetry algebra of
CHS gravity. More precisely, we will show that its total space of
on-shell one-particle states in even dimension $d$ carries the same
(reducible) representation of the conformal algebra $\so(2,d)$ as its
global symmetry algebra (spanning the adjoint representation of the
higher-spin algebra), i.e.
\begin{equation}
  \textrm{On-shell CHS} \quad = \quad \textrm{ Higher-spin Algebra}\,.
  \label{main_identity}
\end{equation} 
Strictly speaking, we will sum the characters of the CHS fields making
up the spectrum of CHS gravity and compare them to the character of
the corresponding conformal higher-spin algebra, which is isomorphic
to the higher-spin algebra of the associated MHS gravity. In such
derivation, it is crucial to use the identity
\begin{equation}
  \chi_{\rm KT}(q,\bm x) = \chi_{\rm CHS}(q,\bm x) + \chi_{\rm
    (P)M}(q, \bm x) - \chi_{\rm (P)M}(q^{-1}, \bm x)\, ,
  \label{rel_KT_FT_PM}
\end{equation}
where $\chi_{\rm KT}$, $\chi_{\rm CHS}$ and $\chi_{\rm{(P)M}}$
designate the $\so(2,d)$ characters of a Killing tensor and its
associated CHS and (partially-)massless field, respectively. Here, $q$
and $\bm x=(x_1,\ldots, x_{\frac d2})$ are related to the temperature
and the chemical potentials for the angular momenta in the usual way
(see \eqref{rel_q_temperature} in the next section).  The identity
\eqref{rel_KT_FT_PM} relates the characters of a given CHS gauge field
on flat spacetime M$_d$, of the associated partially-massless field in
AdS$_{d+1}$ as well as the Killing tensor of the
latter.\,\footnote{Let us recall that any partially-massless field
  $\varphi_\Y$ in AdS$_{d+1}$ with minimal energy $\Delta_{\rm PM}$
  and spin $\Y$ is associated with a conformal field $\phi_\Y$ on
  M$_d$ with conformal weight $d-\Delta_{\rm PM}$ and same spin $\Y$
  (which is used as a source for the dual operator $J_{\Delta,\Y}$ in
  the generating functional of correlation functions of the dual CFT).
  Moreover, the Killing tensors of $\varphi_\Y$ in AdS$_{d+1}$ are
  isomorphic to the conformal Killing tensors of $\phi_\Y$ on M$_d$.
  See \cite{Shaynkman:2004vu, Vasiliev:2009ck} for the analysis in
  full generality, as well as \cite{Metsaev:1999ui, Metsaev:2007rw,
    Metsaev:2009ym, Metsaev:2012hr, Metsaev:2014iwa, Metsaev:2014sfa,
    Chekmenev:2015kzf}.  } This identity was first derived for the
massless totally symmetric case in \cite{Beccaria:2014jxa} (see also
\cite{Giombi:2013yva} for related discussion), then further explored
in \cite{Beccaria:2016tqy}.  The generalized version of the identity
to partially-massless as well as mixed-symmetry cases is derived in
\hyperref[app:zoo]{Appendix \ref{app:zoo}}. The idea of the proof of
\eqref{main_identity} is to perform the sum of \eqref{rel_KT_FT_PM}
over all fields in the spectrum of CHS gravity. After the sum, the
terms corresponding to the last two terms in \eqref{rel_KT_FT_PM}
cancel each other because of parity properties of the corresponding
characters.  A possible physical interpretation of the result
\eqref{main_identity} might be that the dynamical degrees of freedom
of CHS gravity reorganize into its asymptotic symmetries.  In this
sense, CHS gravity is somehow ``topological''. This interpretation
resonates with similar observations on the scattering amplitudes
\cite{Joung:2015eny,Beccaria:2016syk} and the one-loop partition
functions on various backgrounds
\cite{Beccaria:2015vaa,Beccaria:2016tqy} of type-A CHS gravity.

The paper is organized as follows. In \hyperref[sec:typeA]{Section
  \ref{sec:typeA}}, we start by reviewing the field theoretical
definition of type-A CHS theory as spelled out in \cite{Segal:2003jv,
  Bekaert:2010ky}, then move on to the derivation of the property
\eqref{main_identity}. In \hyperref[sec:extension]{Section
  \ref{sec:extension}}, we extend property \eqref{main_identity} to a
couple of classes of higher-depth CHS theories, whose spectrum are
made up of the CHS fields associated to the partially-massless fields
making the spectrum of the type-A$_\ell$ \cite{Bekaert:2013zya} and
type-B$_\ell$ theories. We start by describing the field theoretical
realization of the type-A$_\ell$ CHS theory in
\hyperref[sec:typeAlfield]{Section \ref{sec:typeAlfield}} and move on
to show that \eqref{main_identity} holds for the type-A$_\ell$ in
\hyperref[sec:typeAlchar]{Section \ref{sec:typeAlchar}} and for the
type-B$_\ell$ in \hyperref[sec:typeBl]{Section \ref{sec:typeBl}}.  We
conclude the paper in \hyperref[sec:discu]{Section \ref{sec:discu}} by
discussing our main result and commenting on its implication for the
computation of the thermal partition function and the free energy on
AdS background of this theory. In particular, we point out that
turning on chemical potentials for the $\so(d)$ angular momenta
provides an alternative regularization of the sum over the infinite
tower of higher-spin fields, which we also compare to the previously
used regularization in the literature.  Finally,
\hyperref[sec:typeBl]{Appendix \ref{app:zoo}} contains the derivation
of identity \eqref{rel_KT_FT_PM}, together with a more detailed
description of the various modules of importance and their field
theoretical interpretations in AdS$_{d+1}$/CFT$_d$. In
\hyperref[app:d=2]{Appendix \ref{app:d=2}} we detail the $d=2$ case as
a toy model, while \hyperref[app:branching]{Appendix
  \ref{app:branching}} contains the branching rule of the on-shell
(totally-symmetric) CHS field module.
In \hyperref[app:Segal]{Appendix \ref{app:Segal}}, we provide a short review of nonlinear CHS theory. In particular, we reexamine the formal operator approach of Segal \cite{Segal:2002gd} and provide a heuristic argument supporting the ``topological'' nature of CHS gravity.

\section{Type-A Conformal Higher-Spin Gravity}
\label{sec:typeA}
\subsection{Field theory of type-A conformal higher-spin gravity}
The free theory of the conformal spin-$s$ field in even
$d$-dimensional Minkowski\footnote{Whenever global issues would be
  relevant, M$_d$ should stand for its conformal compactification
  $S^1\times S^{d-1}$.} spacetime M$_d$, is described by the local
action
\begin{equation}\label{actionCHS}
  S_{\rm\sst FT}[h_s] = \int_{M_d} \dd^dx \, h_s\, \proj_{\rm\sst
    TT}^s\, \Box^{s+\frac{d-4}2}\,h_s\, ,
\end{equation}
where $h_s$ is a totally symmetric rank-$s$ tensor and $\proj_{\rm\sst
  TT}^s$ is the projector to transverse and traceless symmetric
tensors of rank $s$.  The differential operator $\Box^{s+\frac{d-4}2}$
compensates the non-locality of $\proj_{\rm\sst TT}^s$ so that the
action is local and conformally invariant.
\begin{itemize}
\item The field $h_s$, referred to as \emph{off-shell
  Fradkin-Tseytlin} (FT) field is a symmetric rank-$s$ field of conformal weight
  $\Delta_{h_s}=2-s$.  It is also called as the \emph{shadow} field.
  Due to the projector $\proj_{\rm\sst TT}^s$, the action has the
  gauge symmetry
  \begin{equation}
    h_s \sim h_s+ \partial \,\xi_{s-1} + \eta\, \sigma_{s-2}\, ,
    \label{gauge tr}
  \end{equation}
  with $\xi_{s-1}$ a rank-$(s-1)$ symmetric tensor, $\eta$ the
  Minkowski metric and $\sigma_{s-2}$ a rank-$(s-2)$ symmetric tensor.
  Here, we used the schematic notation where all the indices are
  implicit.
\item The \emph{conformal Killing tensor} is the set of parameters
  $(\xi_{s-1},\sigma_{s-2})$ satisfying the conformal Killing equation
  \begin{equation}
    \partial \,\xi_{s-1} + \eta\, \sigma_{s-2}=0\,,
    \label{CKT}
  \end{equation}
  i.e. the gauge parameters leaving the FT field inert under
  \eqref{gauge tr}.
\item The equation of motion of the action \eqref{actionCHS} reads
  \begin{equation}
    \proj_{\rm\sst TT}^s\,\Box^{s+\frac{d-4}2}\,h_s\approx 0\,.
    \label{Bach}
  \end{equation}
  This equation is referred to as the spin-$s$ \emph{Bach equation}
  where equalities that only hold on-shell will be denoted by the weak
  equality symbol $\approx$ hereafter.  An equivalence class
  \eqref{gauge tr} of fields $h_s$ obeying this equation will be
  referred to as \emph{on-shell Fradkin-Tseytlin} field.
\item The action \eqref{actionCHS} can be also rewritten, after
  integrating by part, as
  \begin{equation}
    S_{\rm\sst FT}[h_s] = (-1)^s\, \int_{M_d} \dd^dx \ C_{s,s}\,
    \Box^{\frac{d-4}2}\, C_{s,s}\, ,
  \end{equation}
  where $C_{s,s} =\proj_{\rm\sst T}^{s,s}\,\partial^s\,h_s$ is the
  (generalized) \textit{Weyl tensor} of the FT field (here
  $\proj_{\rm\sst T}^{s,s}$ denotes the traceless projector onto the
  two-row Young diagram displayed in \eqref{Weyl} below).  It is a
  traceless tensor with the symmetry of a rectangular two-row Young
  diagram,
  \begin{equation}
    C_{s,s} \quad \underset{\so(1,d-1)}{\sim} \quad {\scriptsize
      \gyoung(_5s,_5s)}\,.
    \label{Weyl}
  \end{equation}
  The Weyl tensor $C_{s,s}$ is a primary field with conformal weight
  $\Delta_{C_{s,s}} = 2$ and is invariant under the gauge
  transformations \eqref{gauge tr}. In particular, for $s=2$ it
  corresponds to the linearized Weyl tensor.
\end{itemize}

So far we have considered the free theory of conformal spin-$s$ gauge
fields. An interacting theory of CHS gauge fields can be constructed
from the effective (also called ``induced'') action of a free scalar
field \cite{Tseytlin:2002gz, Segal:2003jv, Bekaert:2010ky,
  Bonezzi:2017mwr} in a higher-spin background.  Starting from the
action of a free complex scalar field $\phi$ coupled to higher-spin
sources $h_s$ via traceless conserved currents $J_s = \bar\phi\,
\partial^s \phi$ (where $\bar\phi$ denotes the complex conjugate of
$\phi$)\,:
\begin{equation}
  S[\phi; \{h_s\}_{s\in\N}] = \int_{M_d} \dd^dx\, \Big(\bar \phi\,
  \Box\, \phi + \sum_{s=0}^\infty J_s\, h_s\Big)\,,
  \label{noether_int}
\end{equation}
we obtain the effective action
\begin{equation}\label{pathint}
  e^{-W_\Lambda[\{h_s\}_{s\in\N}]} = \int_\Lambda \mathcal D\phi\,
  e^{-S[\phi; \{h_s\}_{s\in\N}]}\,,
\end{equation}
where $\Lambda$ is the ultraviolet (UV) cut-off.  The logarithmically divergent part
$W_{\rm log}$ of the effective action $W_\Lambda$ is a local and
nonlinear functional of the shadow fields $h_s$.  Moreover it can be
shown (see e.g. \cite{Segal:2003jv}) to reproduce the free action
\eqref{actionCHS} at the quadratic order:
\begin{equation}\label{Wlog}
  W_{\log}[\{h_s\}_{s\in\N}] = \sum_{s=0}^\infty S_{\rm\sst FT}[h_s] +
  \mathcal{O}(h_s^3)\,,
\end{equation}
and contains also the interaction terms $\mathcal{O}(h_s^3)$, which
can be perturbatively calculated --- see for instance
\cite{Segal:2003jv,Bekaert:2010ky, Beccaria:2017nco}. Therefore,
$W_{\log}$ can be regarded as an action of interacting FT fields, up
to the introduction of a dimensionless coupling constant $\k$\,:
$S_{\rm\sst CHS}=\k\,W_{\log}$.  Note that the $s=0$ term in
\eqref{Wlog} corresponds to the conformal scalar with $d-4$
derivatives. Hence, for $d=4$ the scalar field becomes auxiliary and
drops out from the spectrum of CHS.

According to the AdS/CFT correspondence, $W_{\rm log}$ should be equal
to its AdS counterpart, that is, the logarithmically divergent part of
the on-shell AdS action in the limit where the location of the
boundary is pushed to infinity.  In this way, the type-A CHS theory is
linked to the type-A MHS theory.  For instance, the free action
$S_{\rm\sst FT}[h_s]$ is related to the Fronsdal action in
AdS$_{d+1}$. Schematically, we have
\begin{equation}
  S_{\rm\sst Fronsdal}[\Phi_s=\mathcal K\,h_s]\,=\, \log
  R\,\,S_{\rm\sst FT}[h_s]\,+\,\textrm{regular or polynomially
    divergent terms},
\end{equation}
where $\mathcal K$ is the boundary-to-bulk propagator of Fronsdal
field $\Phi_s$ and $R$ is the distance from the center to the boundary
of the regularized AdS space (see
e.g. \cite{Mikhailov:2002bp,Metsaev:2008ks,Giombi:2009wh,Joung:2011xb}). The
interacting theory of CHS fields enjoys a non-Abelian global symmetry
generated by the conformal Killing tensors.  From the effective action
point of view, this CHS symmetry is nothing but the maximal symmetry
of the free scalar field.  Hence, it coincides with the type-A MHS
symmetry in AdS$_{d+1}$.

Suppose that we are interested in the quantum properties of CHS theory
such as the one-loop free energy. Then, this quantity is closely
related to the $\so(2,d)$ character of the free CHS theory. The latter
character can be viewed as a single-particle partition function on
$S^1\times S^{d-1}$ (the conformal boundary of thermal AdS$_{d+1}$)
where we turn on, besides the temperature $\beta^{-1}$, the chemical
potentials $\Omega_i$ corresponding to the angular momenta
\cite{Gibbons:2006ij,Dolan:2005wy}, upon the identification,
\begin{equation}\label{rel_q_temperature}
  q = e^{-\beta}\,, \qquad x_i = e^{\beta\,\Omega_i}\,.
\end{equation}
In order to compute the character of the free CHS theory, we need to
determine first the characters of the individual $\so(2,d)$-modules
relevant in CHS theory.

\subsection{Relevant modules}
Let us start by reviewing the relevant modules in the free CHS theory
(see appendix F of \cite{Beccaria:2014jxa}).  As usual in a conformal
field theory, a primary field (together with its descendants) is
described\,\footnote{Strictly speaking, a generalized Verma module is
  the algebraic dual of the infinite jet space at a point of a primary
  field \cite{Eastwood:1987ki}.} by a (generalized) Verma module
$\V\big(\Delta; \Y\big)$, which is induced from the $\so(2) \oplus
\so(d)$ module with lowest weight $[\Delta;\Y]$. Here $\Delta$ is a
real number corresponding to the conformal weight, and
$\Y:=(s_1,\dots,s_r)$ is an integral dominant $\so(d)$-weight, i.e.
\begin{equation}
  \begin{aligned}
    s_1 \geqslant s_2 \geqslant \dots \geqslant s_{r-1} \geqslant
    |s_r|\,, & \qquad [d=2r]\\ s_1 \geqslant s_2 \geqslant \dots
    \geqslant s_r \geqslant 0\quad, & \qquad [d=2r+1]
  \end{aligned}
\end{equation}
where $s_1, \dots, s_r$ are either all integers or all half-integers,
corresponding to the spin, and $r$ is the rank of $\so(d)$. The
irreducible module obtained as a quotient of the Verma module
$\V\big(\Delta; \Y\big)$ by its maximal submodule will be denoted
$\D\big(\Delta; \Y\big)$.

\paragraph{\itshape Rac.}

We introduce first the $\so(2,d)$-module called Rac of order-$\ell$
(or $\ell$-lineton) describing the conformal scalar field with
$\ell$\,th power of the wave operator as kinetic operator. Off-shell,
a scalar field $\phi$ of conformal weight $\frac{d-2\ell}2$ is
described by the Verma module $\V(\frac{d-2\ell}2, \zero)$ where
$\zero$ stands for the trivial $\so(d)$-weight. On-shell, such a
scalar field obeying to the polywave equation,
\begin{equation}
  \Box^\ell\,\phi \approx 0\,,
\end{equation}
is described by the irreducible\,\footnote{Notice that, strictly
  speaking, when $d$ is even the module \eqref{Racell} is irreducible
  if only if $\ell < \frac d2$.} module
\begin{equation}\label{Racell}
  \mbox{Rac}_\ell\,=\,\frac{\V(\frac{d-2\ell}2,
    \zero)}{\V(\frac{d+2\ell}2, \zero)}\,.
\end{equation}
This module is unitarizable for $\ell=1$, in which case it is simply
called ``Rac'' (or scalar singleton). The value $\ell=\frac{d-4}2$ for
the higher-order $\rac$ module gives precisely the $s=0$ part of the
CHS theory spectrum. Notice that the scalar FT field is absent in
$d=4$ and is unitary only for $d=6$, in which case it corresponds to
the usual $\rac$ singleton as $\ell=\frac{d-4}2=1$. When $d \geqslant
8$ however, the order of the scalar singleton is greater than $1$ and
therefore this field is non-unitary, as the on-shell FT fields.

\paragraph{\itshape Conserved current.}

Let us now introduce the module describing the conserved spin-$s$
current $J_s$\,:
\begin{equation}\label{conservation}
  \partial \cdot J_s \approx 0\,.
\end{equation}
i.e. a totally symmetric, traceless and divergenceless rank-$s$
tensor.  This current corresponds to the module $\D\big( s+d-2; (s)
\big)$ defined as the quotient
\begin{equation}
  \D\big( s+d-2; (s) \big) \,=\,\frac{\V\big( s+d-2; (s
   ) \big)}{\V\big( s+d-1; (s-1) \big)}\,.
  \label{cons cur}
\end{equation}
The module $\V\big(s+d-2; (s)\big)$ contains symmetric and traceless
rank-$s$ tensors with conformal weight $\Delta=s+d-2$ whereas the
module $\V\big(s+d-1; (s-1)\big)$ is isomorphic to the divergence of
such tensors. As a consequence, modding out this submodule is
equivalent to imposing the conservation law \eqref{conservation}. From
the AdS$_{d+1}$ perspectives, the \textit{unitarizable} module
\eqref{cons cur} corresponds to the Hilbert space of the massless
spin-$s$ field with Dirichlet boundary conditions, i.e. the
\textit{normalizable} solutions thereof.

\paragraph{\itshape Off-shell Fradkin-Tseytlin field.}

The \textit{off-shell} FT (or shadow) field corresponds to the module
$\Sh\big( 2-s; (s) \big)$ whose field-theoretical realization is a
totally symmetric rank-$s$ tensor field $h_{s}$ quotiented by the
gauge symmetries \eqref{gauge tr}. The case $s=0$ is somewhat
degenerate: the off-shell scalar FT field is simply
$\Sh(2;\zero)=\V(2;\zero)$.  The precise group-theoretical description
of 
this definition of the shadow field in terms of Verma modules remains somewhat
elusive. Notice that this module is not a (quotient of) Verma
module(s) but is rather related to the contragredient thereof. Indeed,
classical field-theoretical terms it is the dual of the conserved
current $J_s$ with respect to the inner product $\int \dd^dx\, h_s\,
J_s$ in the path integral. Equivalently, in CFT terms it is the
algebraic dual of the conserved current $J_s$ with respect to the
identity two-point function $\langle h_s\,
J_{s'}\rangle=\delta_{ss'}$.\footnote{More precisely, it is the
  ``shadow operator'' of $J_s$ in the sense of \cite{Ferrara:1972uq}.}	
	
Fortunately, the equivalent field-theoretical description of the
off-shell FT field $h_s$ in terms of the gauge-invariant spin-$s$ Weyl
tensor $C_{s,s}=\partial^s h_s+\cdots$ has a transparent
group-theoretical description as a submodule of the corresponding
Verma module, i.e. $\Sh\big(2-s;(s)\big)\subset \V\big(2;
(s,s)\big)$. In this case, the previous inner product becomes $\int
\dd^dx\, h_s\, J_s=(-1)^s\int \dd^dx\, C_{s,s}\, k_{s,s}$ where
$k_{s,s}$ is the (non-local) prepotential of the conserved current:
$J_s\approx(\partial\cdot)^sk_{s,s}$. Let us stress that this
description applies to any dimension $d$. However, for $d$ odd, the
module $\Sh\big( 2-s; (s) \big)$ for the off-shell FT field is
irreducible and isomorphic to the module
\begin{equation}
  \D\big(2; (s,s)\big) \cong \frac{\V\big(2; (s,s)\big)}{\D\big(3;
    (s,s,1) \big)}\,,
\end{equation}
describing the spin-$s$ Weyl tensor $C_{s,s}$, where the quotient by
the irreducible submodule $\D\big(3; (s,s,1) \big)$ implements the
generalized Bianchi identities.

\paragraph{\itshape On-shell Fradkin-Tseytlin field.}

For $d$ even, the module $\Sh\big( 2-s; (s) \big)$ of the
\textit{off-shell} FT field is reducible, corresponding to the
possibility of imposing the (local) Bach equation.  Interestingly, the
module \eqref{cons cur} may be interpreted as the left-hand-side of
\eqref{Bach}, i.e. the \textit{Bach tensor}. Accordingly, the module
corresponding to the \textit{on-shell} FT field is in fact given by
the following quotient:
\begin{equation}
  \D\big(2; (s,s)\big) \cong \frac{\Sh\big( 2-s; (s)
    \big)}{\D\big(s+d-2; (s) \big)}\,,
  \label{def_FT_mod}
\end{equation}
where the left-hand-side can be interpreted as the spin-$s$ on-shell
Weyl tensor $C_{s,s}$ given in \eqref{Weyl}. Again $\Sh\big( 2-s; (s)
\big)$ corresponds to the off-shell FT field whereas the quotient by
$\D \big(s+d-2; (s) \big)$ has the interpretation of imposing the Bach
equation \eqref{Bach}. At first glance, it might be curious why the
above quotient module is $\D\big(2; (s,s)\big)$ which has the lowest
weight $2$. While the quotient \eqref{def_FT_mod} corresponds to
  the direct translation of the field-theoretical definition of the
  on-shell FT field, the module $\D\big(2;(s,s)\big)$ can be expressed
  as another quotient using its Bernstein-Gel'fand-Gel'fand (BGG)
sequence (see e.g. \cite{Shaynkman:2004vu}). The latter gives
\begin{equation}
  \D\big(2; (s,s)\big) \cong \frac{\V\big(2; (s,s)\big)}{\U\big(3;
    (s,s,1) \big)}\,,
  \label{FT Weyl_mod}
\end{equation}
where $\U\big(3; (s,s,1) \big)$ implements the identities \`a la
Bianchi obeyed by the on-shell Weyl tensor.  Notice that the
degenerate case $s=0$ is consistent with \eqref{def_FT_mod} in the
sense that the on-shell scalar FT field, which is an order
$\ell=\frac{d-4}2$ scalar singleton, is
\begin{equation}
  \cD(2;\zero) = \frac{\V(2;\zero)}{\V(d-2;\zero)} =
  \rac_{\frac{d-4}2}\,.
\end{equation}

\paragraph{\itshape Conformal Killing tensors.}
The last module we shall introduce is the one corresponding to
conformal Killing tensors,
\begin{equation}
  \D\big( 1-s; (s-1) \big) \cong \frac{\V\big( 1-s; (s-1
   ) \big)}{\U\big( 2-s; (s) \big)}\,.
  \label{KTVerma}
\end{equation}
Here $\V\big(1-s; (s-1)\big)$ corresponds to the (large) gauge
parameters of FT field $h_s$ and $\U\big( 2-s; (s) \big)$ has the
interpretation of pure gauge shadow fields, thus quotienting by this
submodule corresponds to imposing the conformal Killing equation
\eqref{CKT}. Notice that there is a one-to-one correspondence between
Killing tensors on AdS$_{d+1}$ and conformal Killing tensors on M$_d$,
so the finite-dimensional irreducible $\mathfrak{so}(d,2)$-module
$\D\big( 1-s; (s-1) \big)$ has clear bulk and boundary
interpretations.

\subsection{Character of the Fradkin-Tseytlin module}
The characters of the modules presented in the previous section can be
related to the characters of the Verma module,
\begin{equation}
  \chi^{\phantom{g}}_{\cV(\Delta, \Y)}(q, \bm x) = q^\Delta\, \Pd
  d(q,\bm x)\, \chi^{\so(d)}_{\Y}(\bm x)\,,
  \label{V char}
\end{equation}
where $\chi^{\so(d)}_{\Y}$ is the character of the subalgebra $\so(d)$
of $\so(2,d)$ and the function $\Pd d$ is given, both for even and odd
$d$, by
\begin{equation}
  \Pd d (q, \bm x) = \frac{1}{(1-q)^{d-2r}}\, \prod_{k=1}^r
  \frac{1}{(1-q\,x_k)(1-q\,x_k^{-1})}\,.
\end{equation}
 Let us point out one important property of
the above character,
\begin{equation}
  \chi^{\phantom{g}}_{\cV(\Delta,\Y)}(q^{-1},\bm x) = (-1)^d\,
  \chi^{\phantom{g}}_{\cV(d-\Delta,\Y)}(q, \bm x)\,,
  \label{inversion}
\end{equation}
which is simply a consequence of the behaviour of $\Pd d$ under
$q\rightarrow q^{-1}$.

In this paper, we aim to find the character of the total linearized
spectrum of CHS theory.  For that, we need first the character of the
spin-$s$ on-shell FT field --- which had been obtained in
\cite{Beccaria:2014jxa} using the BGG
resolution of finite-dimensional $\so(2,d)$-modules \cite{Shaynkman:2004vu} --- then sum over all
the spins.  In order to avoid the technicalities, we shall present
only key steps of the derivation, but interested readers can find more
details in \hyperref[app:zoo]{Appendix \ref{app:zoo}}.

Let us begin with the character of the spin-$s$ on-shell FT field.
From the definition \eqref{def_FT_mod}, we first find
\begin{equation}
  \chi^{\phantom{g}}_{\cD(2;(s,s))} =
  \chi^{\phantom{g}}_{\Sh(2-s;(s))} -
  \chi^{\phantom{g}}_{\cD(s+d-2;(s))}\,,
  \label{char_shadow}
\end{equation}
where the off-shell FT field character
$\chi^{\phantom{g}}_{\Sh(2-s;(s))}$ and the Bach tensor (or conserved
current) character $\chi^{\phantom{g}}_{\cD(s+d-2;(s))}$ are given by
\begin{eqnarray}
  \chi^{\phantom{g}}_{\Sh(2-s;(s))}\eq
  \chi^{\phantom{g}}_{\cV(2-s;(s))} -
  \chi^{\phantom{g}}_{\cV(1-s;(s-1))} +
	\chi^{\phantom{g}}_{\cD(1-s;(s-1))}\,,
  \label{Sh} \\
  \chi^{\phantom{g}}_{\cD(s+d-2;(s))}\eq \chi^{\phantom{g}}_{\cV(s+d-2;(s))} -
  \chi^{\phantom{g}}_{\cV(s+d-1;(s-1))}\,.
  \label{Cur}
\end{eqnarray}
The equation \eqref{Cur} follows directly from the definition
\eqref{cons cur}.  The heuristic behind \eqref{Sh} is that one should
subtract from the character of the module $\cV(2-s;(s))$ describing
$h_s$ the character of the pure gauge modes. This can be done by
subtracting the character of the module $\cV(1-s;(s-1))$ describing
the gauge parameters. However, this removes too much. Indeed, when the
gauge parameters are equal to conformal Killing tensors, they leave
$h_s$ inert (by definition). For this reason one has to correct by
adding the character of $\cD(1-s;(s-1))$. In more physical terms, the
module $\cV(1-s;(s-1))$ of gauge parameters contains large gauge
transformations --- which are physical --- associated with the
conformal Killing tensor module $\cD(1-s;(s-1))$. Inserting \eqref{Sh}
and \eqref{Cur} in \eqref{char_shadow}, we obtain
\begin{equation}
  \chi^{\phantom{g}}_{\cD(2;(s,s))} =
  \chi^{\phantom{g}}_{\cD(1-s;(s-1))} +
  \chi^{\phantom{g}}_{\cV(2-s;(s))} -
  \chi^{\phantom{g}}_{\cV(1-s;(s-1))} -
  \chi^{\phantom{g}}_{\cV(s+d-2;(s))} +
  \chi^{\phantom{g}}_{\cV(s+d-1;(s-1))}\,,
  \label{relation}
\end{equation}
which relates the spin-$s$ on-shell FT field module
$\chi^{\phantom{g}}_{\cD(2;(s,s))}$ to the character of the conformal
Killing tensor module $\chi^{\phantom{g}}_{\cD(1-s;(s-1))}$ up to the
characters of a few Verma modules. The relation \eqref {relation} does
not yet express the character $\chi^{\phantom{g}}_{\cD(2;(s,s))}$ in
terms of Verma module characters $\chi^{\phantom{g}}_{\cV(\Delta;\Y)}$
alone due to the presence of $\chi^{\phantom{g}}_{\cD(1-s;(s-1))}$ on
the right-hand-side. In principle, we could further work out to get
rid of the latter module using another relation for the modules but it
will turn out to be useful to do the opposite. In fact, the expression
\eqref{relation} naturally leads to an interesting and suggestive
expression of the character of the full CHS theory. This is thanks to
the special property that the Verma module part of \eqref{relation}
enjoys:
\begin{equation}
  \chi^{\phantom{g}}_{\cV(2-s;(s))}(q, \bm x) -
  \chi^{\phantom{g}}_{\cV(1-s;(s-1))}(q, \bm x) =
  (-)^d\chi^{\phantom{g}}_{\cD(s+d-2;(s))}(q^{-1}, \bm x)\,.
  \label{property}
\end{equation}
Note that the above property is a simple consequence of
\eqref{inversion} and \eqref{Cur}. This leads to the relation
\cite{Beccaria:2014jxa}
\begin{equation}
  \chi^{\phantom{g}}_{\Sh(2-s;(s))}(q,\bm x) =
  \chi^{\phantom{g}}_{\cD(1-s;(s-1))}(q,\bm x) +(-)^d
  \chi^{\phantom{g}}_{\cD(s+d-2;(s))}(q^{-1},\bm x)\,,
	\label{off/KT}
\end{equation}
which is valid in any dimension $d>2$ (even or odd). Finally for even
$d$, the character of the spin-$s$ on-shell FT field module coincides
with that of the conformal Killing tensor module up to just two
additional terms:
\begin{equation}
  \chi^{\phantom{g}}_{\cD(2;(s,s))}(q, \bm x) =
  \chi^{\phantom{g}}_{\cD(1-s;(s-1))}(q, \bm x) +
  \chi^{\phantom{g}}_{\cD(s+d-2;(s))}(q^{-1}, \bm x) -
  \chi^{\phantom{g}}_{\cD(s+d-2;(s))}(q, \bm x)\,,
  \label{decompo_CHS_sym}
\end{equation}
as follows from \eqref{char_shadow} and \eqref{property}.
Interestingly, both of these terms are given by the characters of the
conserved current module, but one is with $q$ while the other is with
$q^{-1}$. The formula \eqref{decompo_CHS_sym} is the instance of the
identity \eqref{rel_KT_FT_PM} which is relevant for type-A CHS
gravity. It applies to the degenerate case $s=0$ as well, except that
the first term of the right-hand-side is absent in this case.

\subsection{Character of type-A on-shell conformal higher-spin gravity}
\label{char_on-shell}
We shall use the relation \eqref{decompo_CHS_sym} to derive the
character of the CHS theory linearized spectrum. The theory contains
the FT fields of spin 1 to $\infty$ and the scalar field with a
kinetic operator containing $d-4$ derivatives. Focusing first on the
FT fields, we consider
\begin{eqnarray}
  \sum_{s=0}^\infty\chi^{\phantom{g}}_{\cD(2;(s,s))}(q, \bm x) \eq
  \sum_{s=1}^\infty\chi^{\phantom{g}}_{\cD(1-s;(s-1))}(q, \bm x) + \nn
  && +\, \sum_{s=0}^\infty \chi^{\phantom{g}}_{\cD(s+d-2;(s))}(q^{-1},
  \bm x) - \sum_{s=0}^\infty\chi^{\phantom{g}}_{\cD(s+d-2;(s))}(q, \bm
  x)\,.
  \label{series 1}
\end{eqnarray}
The first term in the right-hand-side of the equality is nothing but
the character of the adjoint module of the CHS symmetry algebra.
Re-expressing the two series in the second line using the
Flato-Fronsdal theorem \cite{Flato:1978qz, Angelopoulos:1999bz,
  Vasiliev:2004cm}:
\begin{equation}
  \Big( \chi^{\phantom{g}}_{\rac}(q, \bm x) \Big)^2 =
  \sum_{s=0}^\infty \chi^{\phantom{g}}_{\cD(s+d-2;(s))}(q, \bm x)\,,
\end{equation}
the series \eqref{series 1} becomes
\begin{eqnarray}
  \sum_{s=0}^\infty\chi^{\phantom{g}}_{\cD(2;(s,s))}(q, \bm x) \eq
  \sum_{s=0}^\infty\chi^{\phantom{g}}_{\cD(1-s;(s-1))}(q, \bm x) + \nn
  && +\, \Big( \chi^{\phantom{g}}_{\rac}(q^{-1}, \bm x) \Big)^2 -
  \Big( \chi^{\phantom{g}}_{\rac}(q, \bm x) \Big)^2\,.
  \label{lines}
\end{eqnarray}
The second line of the above formula vanishes because the character of
the $\rac$ singleton obeys the property
\begin{equation}
  \chi^{\phantom{g}}_{\rac}(q^{-1}, \bm x) = (-1)^{d+1}\,
  \chi^{\phantom{g}}_{\rac}(q, \bm x)\,.
  \label{Racid}
\end{equation}

Finally, we find that the character of all the on-shell fields in the
free CHS theory coincides with that of the global symmetry of CHS
theory:
\begin{equation}
  \sum_{s=0}^\infty \chi^{\phantom{g}}_{\cD(2;(s,s))} =
  \sum_{s=1}^\infty \chi^{\phantom{g}}_{\cD(1-s;(s-1))}\,.
  \label{==}
\end{equation}
This result can be understood as the equality \eqref{main_identity}
for type-A CHS gravity.\footnote{Let us illustrate why
  \eqref{main_identity} does not hold for \textit{minimal} CHS gravity
  with only \textit{even} spins in the spectrum. In such case, the
  Flato-Fronsdal theorem involves a symmetric plethysm of the Rac
  module:
\begin{equation}
  \frac12 \Big[ \Big( \chi^{\phantom{g}}_{\rac}(q,x_i)
    \Big)^2-\chi^{\phantom{g}}_{\rac}(q^2, x_i^2)\Big] = \sum_{s\in
    2{\mathbb N}} \chi^{\phantom{g}}_{\cD(s+d-2;(s))}(q, \bm x)\,.
\end{equation}
Then \eqref{decompo_CHS_sym} and the analogue of \eqref{Racid} imply
the relation
\begin{equation}
  \sum_{s\in 2{\mathbb N}} \chi^{\phantom{g}}_{\cD(2;(s,s))}(q,x_i) =
  \sum_{s\in 2{\mathbb N}_0}
  \chi^{\phantom{g}}_{\cD(1-s;(s-1))}(q,x_i)+\chi^{\phantom{g}}_{\rac}(q^2,
  x_i^2)\,,
\end{equation}
where the extra term on the right-hand-side has no clear
interpretation in this context (as it decomposes into an {\it
  alternating}\, sum of the characters of massless AdS$_{d+1}$ fields
of all integer spin).} Actually, both sides of \eqref{==} involve
divergent series which require some regularization. However, the
equality \eqref{==} itself only assumed the validity of the
Flato-Fronsdal theorem which does not need any
regularization.\footnote{The only assumption is that it holds for each
  sum of characters (in $q$ and in $1/q$) separately, despite the fact
  that the corresponding power series have distinct region of
  convergence ($|q|<1$ and $|q|>1$).} Consequently, confident in the
validity of \eqref{==} one might somehow reduce the issue of
regularizing the character of the CHS spectrum (the left-hand-side) to
the one of the higher-spin algebra\footnote{Some convergence and
  regularization issues of the latter were addressed in
  \cite{Basile:2018dzi}.} (the right-hand-side). By construction, the
corresponding regularization of CHS theory would preserve higher-spin
symmetries, an important requirement of a sensible regularization but
which is usually not guaranteed.

There is another virtuous corollary of the relation
\eqref{main_identity}: the Casimir energy of free CHS theory on the
Einstein static universe ${\mathbb R}\times S^{d-1}$ is ensured to
vanish in any regularization consistent with \eqref{main_identity}. In
fact, the vanishing of the Casimir energy is ensured when the
partition function is invariant under the map $q\to 1/q$ (see
\cite{Giombi:2014yra}) and this property automatically holds for the
right-hand-side of \eqref{main_identity}, since the character of each
Killing tensor module obeys this property. Actually, the $a$-anomaly
of CHS gravity is also guaranteed to vanish by virtue of this property
of the character. Indeed, the $a$-anomaly of a $d$-dimensional FT
field coincides with the difference of the free energy of the
associated massless field in Euclidean AdS$_{d+1}$ with Neumann
boundary condition and the same with Dirichlet condition
\cite{Giombi:2013yva}. Since the free energy with Neumann boundary
condition is simply minus that with Dirichlet condition (the
contribution of Killing tensor module simply vanishes), the
$a$-anomaly of the $d$-dimensional CHS gravity is just minus two times
the free energy of MHS gravity in Euclidean AdS$_{d+1}$. The
cancellation of the latter can be shown using the method of character
integral representation of zeta function
\cite{Bae:2016rgm,Basile:2018zoy,Basile:2018acb}.

\subsection{Character of type-A off-shell conformal higher-spin gravity}
We can also derive an off-shell version of the identity
\eqref{main_identity}. From \eqref{def_FT_mod}, the spin-$s$
off-shell FT field module is related to the on-shell one by
\begin{equation}
  \chi^{\phantom{g}}_{\Sh(2-s;(s))} =
  \chi^{\phantom{g}}_{\cD(2;(s,s))} +
  \chi^{\phantom{g}}_{\cD(s+d-2;(s))}\,,
	\label{off/on}
\end{equation}
whereas the off-shell FT scalar $\Sh(2; 0) =\cV(2;
0)$ is related to the on-shell one by
\begin{equation}
  \chi^{\phantom{g}}_{\Sh(2; 0)}=
  \chi^{\phantom{g}}_{\rac_{\frac{d-4}2}}+
  \chi^{\phantom{g}}_{\cV(d-2; 0)}\,,
\end{equation}
where the first term on the right-hand-side is absent in $d=4$. In
fact, the above off-shell scalar becomes an auxiliary field in four
dimensions. In odd dimension $d$, one can make use of \eqref{off/KT}.
Summing over the characters of these off-shell field modules, we
arrive at
\begin{equation}
  \sum_{s=0}^\infty \chi^{\phantom{g}}_{\Sh(2-s;(s))} =
  \sum_{s=1}^\infty \chi^{\phantom{g}}_{\cD(1-s;(s-1))} +(-)^d
  \sum_{s=0}^\infty \chi^{\phantom{g}}_{\cD(s+d-2;(s))}\,,
  \label{off-shell version}
\end{equation}
which holds for any dimension. For $d$ even, this result can be
viewed as
\begin{equation}
  \mbox{$d$ even:}\qquad \text{Off-Shell CHS} = \text{Higher-spin
    Algebra}\,\oplus\,\text{Dirichlet MHS}\,,
  \label{twAdj_Adj_CHS}
\end{equation}
where the last term on the right-hand-side, the linearized spectrum of
MHS gravity around AdS$_{d+1}$ with Dirichlet boundary conditions,
corresponds to the last series in \eqref{off-shell version}. The two
terms on the right-hand-side of \eqref{twAdj_Adj_CHS} have natural
interpretations in terms of the higher-spin algebra: they are
important modules of the latter. The first term is the adjoint module
while the second term is the so-called twisted adjoint module of the
higher-spin algebra.

From a holographic perspective, another interpretation of this last
result is possible, purely in terms of bulk fields. For a spin-$s$
massless bulk field, two boundary conditions are available: either the
standard (``Dirichlet'') boundary condition which allows normalizable
bulk solutions corresponding to a conserved current $J_s$ with
conformal weight $\Delta_+=s+d-2$, or the exotic (``Neumann'')
boundary condition which allows non-normalizable bulk solutions
corresponding to a shadow field $h_s$ with conformal weight
$\Delta_-=2-s$. Accordingly, the bulk theory with standard
(respectively, exotic) boundary condition for \textit{all} fields will
be referred to as Dirichlet (respectively, Neumann) MHS theory. They
are summarized and compared in \hyperref[list]{Table \ref{list}}. The
holographic dual of Dirichlet MHS theory is a free scalar CFT
\cite{Sezgin:2002rt, Klebanov:2002ja}. Following the usual
considerations on double-trace deformations and holographic degeneracy
\cite{Klebanov:1999tb, Witten:2001ua, Sever:2002fk}, applied for all
spin-$s$ conserved currents, one is lead to the conclusion that the
holographic dual of Neumann MHS theory is CHS gravity. This scenario
has been extensively discussed for the MHS theory around AdS$_4$ and
Chern-Simons CHS theory around M$_3$ (see e.g. \cite{Leigh:2003ez,
  Vasiliev:2012vf, Giombi:2013yva}), but the logic works for any
dimension (see e.g. \cite{Metsaev:2009ym} for any spin at free level
and \cite{Compere:2008us} for spin-two at interacting level).
	
\begin{table}[h]
\centering
\begin{tabular}{|c|c|c|c|c|}
 \hline
Bulk MHS & Bulk & $\Delta_s$ & Boundary & Boundary 
 \\ 
theory & field & & operator 
 & theory 
\\ \hline\hline
Dirichlet & Normalizable & $\Delta_+=$ & Conserved & Free 
\\
(standard) & solution & $s+d-2$ & current &  CFT 
 \\\hline
Neumann & Non-normalizable & $\Delta_-=$ & Shadow  & CHS 
\\
(exotic) & solution & $2-s$ & field  & theory 
 \\\hline
\end{tabular}
\caption{List of relevant fields in Dirichlet vs Neumann MHS theories}
\label{list}
\end{table}

This exotic type of holographic dualities where both sides can be
gravity theories (though of Einstein vs Weyl type) has been denoted
``AdS/IGT'' in \cite{Giombi:2013yva} -- where IGT stands for induced
gauge theory -- in order to distinguish it from standard AdS/CFT
correspondence. This type of holographic duality is somewhat less
familiar and is a subtle one, so let us expand a little bit and
present some details. One starts by adding double-trace deformations
for all $U(N)$-singlet primary operators $J_s$ (with $s=0,1,2,\cdots$)
bilinear in the $N$ complex scalar $\phi^i$:
\begin{equation}
  S_{\{\lambda_s\}_{s\in\N}}[\{\phi^i\}_{i=1,\cdots, N},
    \{h_s\}_{s\in\N}]\, =\, S[\{\phi^i\}_{i=1,\cdots,
      N},\{h_s\}_{s\in\N}]\, +\, \sum_{s=0}^\infty\,
  \frac{\lambda_s}{2N} \int \dd^dx\, J_s^2\,.
\end{equation}
These deformations explicitly break all gauge symmetries of the
background fields $h_s$.  Notice that essentially all\footnote{Except
  possibly for very low spin $s$ and dimension $d$ which require a
  separate discussion (see e.g. \cite{Bekaert:2012ux} for detailed
  discussion of the double-trace deformation and its holographic
  interpretation in the $s=0$ case). More precisely, for $d=4$ the
  $s=0$ term is marginal while for $d=2$ the $s=0$ term is relevant
  and the $s=1$ is marginal. \label{foot}} these deformations are
irrelevant in the infrared (IR).  The corresponding effective action
is defined as
\begin{equation}
  e^{-N\,W^{\{\lambda_s\}_{s\in\N}}_\Lambda[\{h_s\}_{s\in\N}]} =
  \int_\Lambda \bigg[\prod_{i=1}^N \mathcal D\phi^i\bigg]\,
  e^{-S_{\{\lambda_s\}_{s\in\N}}[\{\phi^i\}_{i=1,\cdots,
        N},\{h_s\}_{s\in\N}]}\,.
  \label{connfull}
\end{equation}
The standard Hubbard-Stratonovich trick corresponds to the
introduction of auxiliary fields $\sigma_s$ through Gaussian integrals
as follows:
\begin{equation}
  e^{-\frac{\lambda_s}{2N} \int \dd^dx\, J_s^2} = \int
  \mathcal{D}\sigma\, e^{\int \dd^dx\, \Big((\sigma_s-h_s)\, J_s\, -\,
    \frac{N}{2\lambda_s}\, (\sigma_s-h_s)^2\Big)}\,,
\end{equation}
where we neglected an infinite prefactor. Note that $\sigma_s$ does
not have any gauge symmetries. The integration over the dynamical
scalar field $\phi$ is now again a Gaussian integral which can now be
performed leading to
\begin{equation}
  e^{-N\,W^{\{\lambda_s\}_{s\in\N}}_\Lambda[\{h_s\}_{s\in\N}]} =
  \int_\Lambda \mathcal{D}\sigma\,
  e^{-N\,\Big(\,W_\Lambda[\{\sigma_s\}_{s\in\N}]\, -\,
    \sum\limits_{s=0}^\infty\frac{1}{2\lambda_s} \int \dd^dx\,
    (\sigma_s-h_s)^2\,\Big)}\,.
    \label{HStransform}
\end{equation}
One may then perform the field redefinition $h_s=\lambda_s\,j_s$ such
that the new background field $j_s$ has the bare scaling dimension of
a conserved current,
\begin{equation}
  e^{-N\,W^{\{\lambda_s\}_{s\in\N}}_\Lambda[\{\lambda_s\,j_s\}_{s\in\N}]}
  = e^{N\,\sum\limits_{s=0}^\infty\frac{\lambda_s}{2} \int \dd^dx\,
    j^{2}_s} \int_\Lambda \mathcal{D}\sigma\,
  e^{-N\,\Big(\,W_\Lambda[\{\sigma_s\}_{s\in\N}]\, +\,
    \sum\limits_{s=0}^\infty \int \dd^dx\, \big(\sigma_s\,j_s-
    \frac{1}{2\lambda_s}\, \sigma^2_s\big)\,\Big)}\,.
  \label{HStransformed}
\end{equation}
Considering the vicinity of the UV fixed point where\footnote{Except
  possibly for low spin $s$ and dimension $d$ which might require a
  separate discussion which will be avoided here for the sake of
  simplicity.} $\lambda_s\to\infty$, one gets
\begin{equation}
    e^{-N\,W_\Lambda^{\rm UV}[\{j_s\}_{s\in\N}]} = \int_\Lambda
    \mathcal{D}h\, e^{-N\,\Big(\,W_\Lambda[\{h_s\}_{s\in\N}]\, +\,
      \sum\limits_{s=0}^\infty h_s\,j_s\,\Big)}\,,
    \label{HStransformedd}
\end{equation}
where $W_\Lambda^{\rm UV}[\{j_s\}_{s\in\N}]$ stands for the
UV-divergent part of the functional
$W^{\{\lambda_s\}_{s\in\N}}_\Lambda[\{\lambda_s\, j_s\}_{s\in\N}] +
\sum_{s=0}^\infty\frac{\lambda_s}{2}\int \dd^dx\, j^{2}_s$\,. In
\eqref{HStransformedd}, the Hubbard-Stratonovitch field $\sigma_s$ has
been denoted $h_s$ (consistently with its bare scaling dimension) in
order to stress that gauge symmetry is restored in the limit
$\lambda_s\to\infty$ due to the disappearance of the quadratic
term.\footnote{Therefore \eqref{HStransformedd} would require a
  careful discussion of the the ghost contribution in the
  measure. This will not be performed here because we only intend to
  sketch the logic of the proof.} Focusing on the logarithmically
divergent piece, the left-hand-side in \eqref{HStransformedd} can be
interpreted as the generating functional of the correlators in CHS
gravity.  In particular, in the large-$N$ (i.e. semiclassical) limit
one has that the logarithmically divergent piece of $W_\Lambda^{\rm
  UV}[\{j_s\}_{s\in\N}]$ is the Legendre transform of $W_{\rm
  log}[\{h_s\}_{s\in\N}]$.

The holographic dual to a Legendre transform on the boundary is a
change of boundary conditions from Dirichlet to Neumann on the bulk
fields. In fact, the solution space of the spin-$s$ MHS fields with
Neumann boundary condition is an $\mathfrak{so}(2,d)$ module in
one-to-one correspondence with the module of a spin-$s$ shadow field
(see e.g. \cite{Bekaert:2012vt} for a manifestly conformal and gauge
invariant field-theoretical description).  Consequently, the
linearized spectra of off-shell CHS theory and Neumann MHS are in
one-to-one correspondence. Therefore the result \eqref{twAdj_Adj_CHS}
can be rephrased purely in bulk terms as follows:
\begin{equation}
  \mbox{$d$ even:}\qquad \text{Neumann MHS} = \text{Higher-spin
    Algebra}\,\oplus\,\text{Dirichlet MHS}\,.
  \label{NMHS}
\end{equation}
In other words, our character computation suggests that asymptotic
charges account for all extra dynamical degrees of freedom in MHS
theory when all boundary conditions are modified from Dirichlet to
Neumann ones. Let us stress that both holographic duals to Dirichlet
and Neumann MHS theories are (respectively, infrared and ultraviolet) fixed points
with \textit{unbroken} conformal higher-spin symmetries (rigid
symmetries in the former case, gauge symmetries in the latter) since
all spins are on the same footing.

As a side remark, one may observe that, the opposite sign in the last
term on the right-hand-side of \eqref{off/KT} for $d$ odd leads after
summation over all spins to a relation between characters which one
can rewrite as
\begin{equation}
  \mbox{$d$ odd:}\quad \text{Dirichlet MHS}\oplus\text{Neumann MHS} =
  \text{Higher-spin Algebra}\,\,.
\end{equation}
This relation suggests that the linearized MHS theory on AdS$_{d+1}$
spacetime of even dimension \textit{without} imposing any boundary
condition (i.e. considering both normalizable \textit{and}
non-normalizable solutions) might also be somewhat ``topological'' in
the sense that its dynamical degrees of freedom reorganize into its
asymptotic symmetries.

\section{Extensions to Type-A$_\ell$ and Type-B$_\ell$ Theories}
\label{sec:extension}
In the previous section, we have shown that the character of the
type-A CHS theory coincides with the character of the adjoint module
of the type-A higher-spin algebra.  In this section, we provide more
non-trivial evidences of this intriguing observation by generalizing
the result to the type-A$_\ell$ and type-B$_\ell$ theories.

\subsection{Field theory of type-A$_\ell$ theory}
\label{sec:typeAlfield}

Let us introduce another class of conformal gauge fields which are
cousins of the spin-$s$ FT field. They are described by
totally-symmetric rank-$s$ tensor $h_s^{(t)}$ like the usual FT field,
but they have weaker gauge symmetry \cite{Bekaert:2013zya}
\begin{equation}
  \delta_{\xi, \sigma} h^{\sst (t)}_{s} = \partial^t\, \xi_{s-t} +
  \eta\, \sigma_{s-2}\, ,
  \label{t CK}
\end{equation}
with $\xi_{s-t}$ a rank-$(s-t)$ symmetric tensor, $\eta$ the Minkowski
metric and $\sigma_{s-2}$ a rank-$(s-2)$ symmetric tensor. The integer
$t$ takes value inside the range $1\leqslant t \leqslant s$,
parameterizes these class of fields and will be referred to as the
\emph{depth} (so that the usual FT field corresponds to $t=1$). As in
the usual FT field case, we can define a gauge-invariant
field-strength, that is, a Weyl-like tensor as
\begin{equation}
  C^{\sst (t)}_{s,s-t+1} = \proj_{\rm\sst
    T}^{s,s-t+1}\,\partial^{s-t+1}\,h^{\sst (t)}_{s}
  \underset{\so(1,d-1)}{\sim} \quad {\scriptsize
    \gyoung(_6s,_5{s-t+1})}\,,
\end{equation}
which also has conformal weight $\Delta_{C^{\sst (t)}_{s,s-t+1}} =
2$\,. In even $d$ dimensions, the action for the depth-$t$ and
spin-$s$ FT field is then given by
\begin{equation}
  S_{\sst{\rm FT}^{\sst (t)}}[h^{\sst (t)}_{s}] = (-1)^{s-t+1}\,
  \int_{M_d} \dd^dx \ C^{\sst (t)}_{s,s-t+1}\, \Box^{\frac{d-4}2}\,
  C^{\sst (t)}_{s,s-t+1}\,.
\end{equation}
After integrating by part, the action takes the form of
\begin{equation}
  S_{\sst{\rm FT}^{\sst (t)}}[h^{\sst (t)}_{s}] = \int_{M_d} \dd^dx
  \ h^{\sst (t)}_{s}\,\proj^{s}_{\sst{\rm T}^{t}\rm
    T}\,\Box^{s-t+\frac{d-2}2} h^{\sst (t)}_{s}\,,
  \label{actionCHSAl}
\end{equation}
where $\proj^{s}_{\sst{\rm T}^{t}\rm T}$ is the $t$-ple transverse and
traceless projector\,\footnote{Note that the condition of the $t$-ple
  transversality and tracelessness does not fix the projector
  uniquely. We need to impose the locality condition on
  $\proj^{s}_{\sst{\rm T}^{t}\rm T}\,\Box^{s-t+1}$ to determine the
  action uniquely.} which becomes local after multiplying by the
factor $\Box^{s-t+1}$.  The condition $\delta_{\xi, \sigma} h^{\sst
  (t)}_{s} =0$ defines now the depth-$t$ conformal Killing tensors.

An interacting theory of the depth-$t$ FT fields can be obtained as an
effective action, similarly to the $t=1$ case.  We replace the free
scalar action by its analog of order $2\ell$ in the derivatives and
couple the system to the higher-spin sources $h^{\sst (t)}_s$ via a
set of currents $J^{\sst (t)}_s$ which are traceless and $t$-ple
divergenceless:
\begin{equation}
  \eta\cdot J^{\sst (t)}_s\approx 0\,, \qquad (\partial\cdot)^t
  J^{\sst (t)}_s\approx 0\,.
  \label{JJ}
\end{equation}
One can show that for a given free scalar action with a fixed $\ell$,
we can find currents of all integers spin with $t=1, 3, \ldots,
2\ell-1$\, \cite{Bekaert:2013zya, Brust:2016gjy, Gliozzi:2017hni}. These currents take
the form
\begin{equation}
  J^{\sst (2k-1)}_s = \bar\phi\, \partial^s\, \Box^{\ell-k}\, \phi +
  \cdots\,, \qquad (k=1,2,\dots,\ell)
\end{equation}
where the ``\dots'' stands for additional terms ensuring \eqref{JJ},
and it has the conformal weight
\begin{equation}
  \Delta_{s,k}=s+d-2k\,.
\end{equation}
The tensors $J^{\sst (t)}_s$ with $t>s$ do not satisfy any
(partial-)conservation condition since \eqref{JJ} is not defined in
such case.  Still, these tensors can be used as part of the basis
operators for the space of operators bilinear in the order-$\ell$
scalar singleton.  When $\ell\leqslant \frac d4$, all these operators
are primary, and the space of operators with dimension $s+d-2k$ and
spin $s$ is spanned by the basis
\begin{equation}
  \{ J_s^{\sst (2k-1)}, \Box\,J_s^{\sst (2k+1)}, \ldots,
  \Box^\ell\,J_s^{\sst (2k+2\ell-1)}\}\,.
  \label{basis}
\end{equation}

Starting from the $2\ell$-derivative scalar field action in the
background of higher-spin fields of depths $t=1,3,\dots,2\ell-1$,
\begin{equation}
  S\big[\phi;\{h_s^{\sst (2k-1)}\}_{s\in\N\,, \, k \in
      \{1,2,\dots,\ell\}}\big] = \int_{M_d} \dd^dx\, \Big(\bar\phi\,
  \Box^\ell\, \phi + \sum_{s=0}^\infty \sum_{k=1}^\ell J^{\sst
    (2k-1)}_s\, h^{\sst (2k-1)}_s\Big)\,,
  \label{noether_int ell}
\end{equation}
and integrating out the scalar field $\phi$, we obtain an effective
action. Again, the logarithmically divergent part of the effective
action is a local functional of higher-spin fields, and for
$\ell\leqslant \frac{d}4$, it has the structure
\begin{equation}
  W^{\sst (\ell)}_{\log}\big[\{h_s^{\sst(2k-1)}\}_{s\in\N\,, \, k \in
      \{1,2,\dots,\ell\}}\big] = \sum_{s=0}^\infty \sum_{k=1}^\ell
  S^{\sst (2k-1)}_{\rm\sst FT}[h^{\sst (2k-1)}_s] +
  \mathcal{O}(h^3)\,.
  \label{ell W}
\end{equation}
The functional $W^{\sst (\ell)}_{\log}$ can be regarded as an action
of interacting higher-depth FT fields up to the introduction of a
dimensionless coupling constant $\k$\,: $S^{\sst (\ell)}_{\rm\sst CHS}
= \k\, W^{\sst (\ell)}_{\log}$.  This interacting theory contains not
only the higher-depth FT fields but also other non-gauge conformal
fields, referred to as \emph{special} in \cite{Metsaev:2016oic}. We
will refer to this class of fields as ``special FT'' for the sake of
uniformity in the terminology. They correspond to the fields of
spin-$s$ and conformal weight $\Delta=1-s+t$ with $t \geqslant
s+1$. Although they do not enjoy any gauge symmetry, we will keep
referring to the parameter $t$ defining those fields as their
depth. In the quadratic part \eqref{ell W}, the fields with $0
\leqslant s \leqslant 2\,(\ell-1)$ and $\frac{s+2}2 \leqslant
k\leqslant \ell$ correspond to this class.  The free Lagrangians of
these fields still has $2(s-t)+d-2$ derivatives but do not have any
gauge symmetry. A trivial but important restriction to these fields is
that the number of derivatives of their free Lagrangian cannot be
negative. This gives a dimension dependent upper bound for $t$, namely
$t\leqslant s + \frac{d-2}2$, and hence $k \leqslant
\frac{d+2s}4$. The latter bound is irrelevant when it is not smaller
than $\ell$. The $s=0$ case gives the lowest bound, and hence the
condition that this be not smaller than $\ell$ imposes $\ell \leqslant
\frac{d}4$\,. From the CFT point of view, the bound implies that there
is no operators with conformal weights lower than $\frac{d}2$.
This bound can actually be relaxed without encountering an
inconsistency due to a subtle phenomenon discussed below.  The origin
of the quadratic Lagrangian in \eqref{ell W} is the local contact
terms hidden in the two point functions of $J_s^{\sst (2k-1)}$\,:
\begin{equation}
  \la J_s^{\sst (2k-1)}(x)\,J_s^{\sst (2k-1)}(0) \ra \propto
  \frac{\eta^s}{|x|^{2s+2d-4k+\epsilon}} + \cdots \underset{\epsilon
    \sim 0}{\longrightarrow} \frac{\rho_{\Delta_{s,k}}}{\e} \left(
  \eta^s\, \Box^{s+\frac{d}2-2k} + \cdots \right) \delta^{\sst
    (d)}(x)\,,
\end{equation}
where $\rho_{\Delta} = \pi^{\frac d2}/[2^{2\Delta-d}\, \Gamma(\Delta -
  \frac{d-2}2)\, \Gamma(\Delta)].$ One can note here that the contact
terms are absent for $k>\frac{d+2s}4$.

This higher-depth CHS theory appears from the on-shell action of
type-A$_\ell$ MHS theory in AdS$_{d+1}$. Analogously to the usual
FT/massless case, the depth-$t$ FT fields in M$_d$ can be related to
the depth-$t$ partially-massless field in AdS$_{d+1}$:
\begin{equation}
  S_{\sst{\rm PM}^{(t)}}[\Phi^{\sst (t)}_s = \mathcal K^{\sst
      (t)}\,h^{\sst (t)}_s] = \log R\,S_{\sst{\rm FT}^{\sst
      (t)}}[h^{\sst (t)}_s] + (\textrm{regular or polynomially
    divergent terms}).
\end{equation}
Here, $\Phi^{\sst (t)}_s$ is the spin-$s$ and depth-$t$
partially-massless field and $\mathcal K^{\sst (t)}$ is the
corresponding boundary-to-bulk propagator. Both type-A$_\ell$
theories (CHS gravity in M$_d$ and partially-massless HS gravity
in AdS$_{d+1}$) have the same global symmetries: the type-A$_\ell$ HS
symmetry algebra generated by higher-depth conformal Killing tensors.

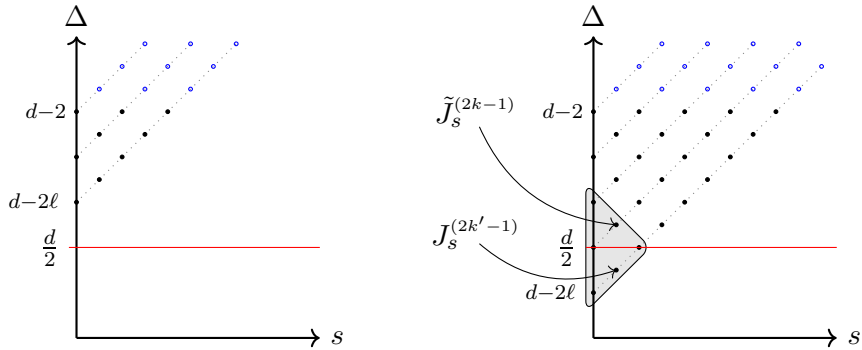
\begin{figure}[!h]
  \centering
  \begin{subfigure}{.4\textwidth}
  \centering
  \begin{tikzpicture}
    \draw[thick,->]  (0,0) -- (3.2,0) node[anchor=west] {$s$};
    \draw[thick,->] (0,0) -- (0,4) node[anchor=south] {$\Delta$};
     \draw[thin,gray,dotted]  (0,1.8) -- (2.1,3.9);
          \draw[thin,gray,dotted]  (0,2.4) -- (1.5,3.9);
     \draw[thin,gray,dotted]  (0,3) -- (0.9,3.9);
       \foreach \y in {0,...,2}{
        \foreach \x in {-\y,...,\y}{
        \node[draw,circle,fill,inner sep=0.5pt] at (0.3*\x+0.3*\y,3+0.3*\x-0.3*\y) {};}};
        \foreach \y in {1,...,3}{
	\foreach \x in {0,...,2}{
        \node[draw,circle,blue, inner sep=0.5pt] at (0.6*\x+0.3*\y,3+0.3*\y) {};}};
	\node[anchor=east] at (0,3) {$\scriptstyle d-2$};
        \node[anchor=east] at (0, 1.8) {$\scriptstyle d-2\ell \,\,$};
	\draw[red] (3.2,1.2) -- (-0.1,1.2) node[black, anchor=east] {$\frac{d}2$};
      \end{tikzpicture}
  \end{subfigure}
   \begin{subfigure}{.4\textwidth}
  \centering
    \begin{tikzpicture}
    \draw [rounded corners,fill=gray!20] (-0.1,0.35)--(0.75,1.2)--(-0.1,2.05)--cycle;
          \draw[thin,gray,dotted]  (0,0.6) -- (3,3.6);
        \draw[thin,gray,dotted]  (0,1.2) -- (2.7,3.9);
        \draw[thin,gray,dotted]  (0,1.8) -- (2.1,3.9);
          \draw[thin,gray,dotted]  (0,2.4) -- (1.5,3.9);
     \draw[thin,gray,dotted]  (0,3) -- (0.9,3.9);
    \draw[thick,->]  (0,0) -- (3.2,0) node[anchor=west] {$s$};
    \draw[thick,->] (0,0) -- (0,4) node[anchor=south] {$\Delta$};
        \foreach \y in {0,...,4}{
        \foreach \x in {-\y,...,\y}{
        \node[draw,circle,fill,inner sep=0.5pt] at (0.3*\x+0.3*\y,3+0.3*\x-0.3*\y) {};}};
        \foreach \y in {1,...,2}{
	\foreach \x in {0,...,4}{
        \node[draw,circle,blue, inner sep=0.5pt] at (0.6*\x+0.3*\y,3+0.3*\y) {};}};
        \foreach \x in {0,...,3}{
         \node[draw,circle,blue, inner sep=0.5pt] at (0.6*\x+0.9,3.9) {};};
	\node[anchor=east] at (0,3) {$\scriptstyle d-2$};
        \node[anchor=east] at (0, 0.6) {$\scriptstyle d-2\ell \,\,$};
	\draw[red] (3.2,1.2) -- (-0.1,1.2) node[black, anchor=east] {$\frac{d}2$};
	\path[<-] (0.3,0.9) edge [bend left] (-1.5,1.25);
	\node at (-1.55,1.4) {$J_s^{\sst (2k'-1)}$};
	\path[<-] (0.3,1.5) edge [bend left] (-1.5,2.8);
	\node at (-1.55,3) {$\tilde J_s^{\sst (2k-1)}$};
      \end{tikzpicture}
  \end{subfigure}
  \caption{The spectrum of bilinear operators, labeled by $(\Delta,
    s)\,$, of the order-$\ell$ scalar CFT is depicted.  The black
    solid circles and the blue empty circles designate respectively
    the operators without conservation condition and with
    (partial-)conservation condition.  The number of the dotted lines
    is $\ell$. The left diagram is the case with $\ell<\frac{d}4$,
    whereas the right one with $\frac{d}4\leqslant \ell <\frac{d}2$.
    The operators in the shaded region do not give rise to on-shell FT
    fields.}
    \label{fig}
\end{figure}

When $\ell>\frac d4$, we face an interesting phenomenon referred to as
\textit{extension} on the CFT side in \cite{Brust:2016gjy}. Let
us briefly review this extension hereafter. For $\ell>\frac d4$, the
operator spectrum contains pairs of currents $J_s^{\sst (2k-1)}$ and
$J_s^{\sst (2k'-1)}$ with the same spin but with the respectively dual
conformal weights, $\Delta_{s,k}+\Delta_{s,k'}=d$, in other words,
\begin{equation}
  k + k' = s + \frac{d}2\,, \qquad k \leqslant k'\,.
  \label{k cond}
\end{equation}
For $\ell< \frac d 2$, none of these currents are
(partially-)conserved (see Fig.\ref{fig}), and the current operator with higher conformal
dimensions becomes a descendent of the other:
\begin{equation}
  J_s^{\sst (2k-1)} \propto \Box^{k'-k}\, J_s^{\sst (2k'-1)}\,.
  \label{degen}
\end{equation}
Hence, the operators $J_s^{\sst (2k-1)}$ are both primary and
descendent. Because of the degeneracy \eqref{degen}, we loose one
basis operator from \eqref{basis}. As a consequence, we can find a new
basis operator $\tilde J_s^{\sst (2k-1)}$ which is neither primary nor
descendent.\,\footnote{Notice that this phenomenon is a
    consequence of the fact that the tensor product of two
    order-$\ell$ scalar singletons contains reducible (though indecomposable)
    representations for $\ell>\frac d4$.}  Interestingly, the cross
two point function of $J_s^{\sst (2k'-1)}$ and $\tilde J_s^{\sst
  (2k-1)}$ does not vanish but gives
\begin{equation}
  \la J_s^{\sst (2k'-1)}(x)\,\tilde J_s^{\sst (2k-1)}(0)\ra \propto
  \frac{\eta^s}{|x|^{d+\e}} + \cdots \underset{\epsilon \sim
    0}{\longrightarrow} \frac{\rho_{\frac d2}}{\e}\, \eta^s\,
  \delta^{\sst (d)}(x)\,.
\end{equation}
This implies that the scalar field action \label{noether_int ell} in
higher-spin background, where all currents $J_s^{\sst (2k-1)}$ with
the condition \eqref{k cond} are replaced by $\tilde J_s^{\sst
  (2k-1)}$ will lead to the effective action $W^{\sst (\ell)}_{\rm
  log}$ containing the quadratic terms
\begin{equation}
  \tilde h_s^{\sst (2k-1)} \left( \Box^{s+\frac{d}2-2k} + \cdots
  \right)\tilde h_s^{\sst (2k-1)} + \tilde h_s^{\sst (2k-1)}\, h_s^{\sst (2k'-1)}\,,
\end{equation}
where $\tilde h_s^{\sst (2k-1)}$ are the source fields of the
operators $\tilde J_s^{\sst (2k-1)}$ and we dropped numerical
coefficients in each terms. This shows that the FT fields $h_s^{\sst
  (2k'-1)}$ with $k'>\frac{d+2s}4$ are in fact Lagrange multipliers
enforcing $\tilde h_s^{\sst (2k-1)}\approx 0$\,. Hence, both of
$h_s^{\sst (2k'-1)}$ and $\tilde h_s^{\sst (2k-1)}$ with \eqref{k
  cond} are absent in the on-shell spectrum. It is shown in
\cite{Brust:2016zns} that the AdS counterpart of this extension
phenomenon is the mass term mixing between the fields $\Phi^{\sst
  (2k-1)}_s$ and $\Phi^{\sst (2k'-1)}_s$\,.

Since the operators $J_s^{\sst (2k'-1)}$ with \eqref{k cond} have
conformal weights lower than $\frac d2$, we can think of deforming
the $U(N)$ free CFT with the double trace deformations of such
operators:
\begin{equation}
  \int_{M_d} \dd^dx\,\bigg(\bar\phi^a\,\Box^\ell\,\phi_a + \sum_{s,
    k'} g_{s,k}\,J_s^{\sst (2k'-1)}\,J_s^{\sst (2k'-1)}\bigg)\,.
  \label{dt def}
\end{equation}
This will lead to an interacting CFT in IR, where a new class of
operators $\cJ_s^{\sst (2k'-1)}$ having the same conformal weight
as $J_s^{\sst (2k-1)}$ may arise. From the AdS point of view, the
double-trace deformation corresponds to changing the boundary
conditions of the relevant fields. It will be interesting to clarify
how the ``extension'' phenomenon will affect the mechanism of
double-trace deformation/changing boundary conditions.

When $\ell\geqslant \frac{d}2$, the solution space of
$\Box^\ell\,\phi\approx 0$ contains a conformally-invariant subspace.
Hence, we are lead to consider either the finite-dimensional quotient
part, or the subspace part. In fact, the quotient space corresponds to
the space of harmonic polynomials with maximum order $\ell-\frac
d2$\,. The zero mode of the $\ell=1$ and $d=2$ case is the familiar
example. The CFT based on the quotient part has been studied in
\cite{Brust:2016gjy}.  We postpone the analysis of CHS theory in this
case to a future work.

\subsection{Character of type-A$_\ell$ theory}
\label{sec:typeAlchar}
The modules relevant to the type-A$_\ell$ CHS theory are analogous to
those of the $\ell=1$ theory: they are
\begin{itemize}
\item the module $\cD(s-t+d-1;(s))$ of $t$-ple conserved current of
  spin-$s$,
\item the modules $\cD(2,(s,s-t))$ and $\cS(1+t-s,(s))$ of the on- and
  off-shell FT fields of spin-$s$ and depth-$t$ (introduced in
  \cite{Bekaert:2013zya}), and
\item the module $\cD(1-s;(s-t))$ of the conformal Killing tensor
 which is a finite-dimensional irrep
    corresponding to the $\so(2,d)$ two-row Young diagram
  ${\footnotesize \gyoung(_5{s-1},_4{s-t})}$\ \cite{Nikitin1991,
    Nikitin2005}.
\end{itemize}
The character of the partially-conserved current module is simple:
\begin{equation}
  \chi^{\phantom{g}}_{\cD(s+d-t-1;(s))}(q, \bm x) = q^{s+d-t-1}\,
  \Big( \chi^{\so(d)}_{(s)}(\bm x) - q^t\, \chi^{\so(d)}_{(s-t)}(\bm
  x) \Big) \Pd d (q, \bm x)\, ,
  \label{character_PM_sym}
\end{equation}
whereas the characters of the other modules are more involved and
their explicit forms are given in \hyperref[app:zoo]{Appendix
  \ref{app:zoo}}. What is important for our purpose is that these
characters satisfy an analogous relation to \eqref{decompo_CHS_sym},
\begin{equation}
  \chi^{\phantom{g}}_{\cD(2;(s,s-t+1))}(q,\bm x) =
  \chi^{\phantom{g}}_{\cD(1-s;(s-t))}(q,\bm x) +
  \chi^{\phantom{g}}_{\cD(s+d-t-1;(s))}(q^{-1}, \bm x)-
  \chi^{\phantom{g}}_{\cD(s+d-t-1;(s))}(q, \bm x)\, .
  \label{decompo_CHS_PM}
\end{equation}
Using the above properties together with the type-A$_\ell$
Flato-Fronsdal theorem \cite{Bekaert:2013zya, Basile:2014wua}
\begin{equation}
  \left(\chi^{\phantom{g}}_{\rac_\ell}\right)^2 =
  \sum_{t=1,3,\dots}^{2\ell-1} \left[\sum_{s=0}^{t-1}
    \chi^{\phantom{g}}_{\cV(s-t+d-1;(s))} + \sum_{s=t}^\infty
    \chi^{\phantom{g}}_{\cD(s-t+d-1;(s))} \right],
  \label{FF_l}
\end{equation}
we can handily compute the character of the entire type-A$_\ell$ CHS
theory. Here, the module $\cV(s-t+d-1;(s))$ corresponds to a spin-$s$
operator without any conservation condition. The field content of the
type-A$_\ell$ CHS theory consists of
\begin{itemize}
\item the FT fields with depth $t=1,3,\ldots, 2\ell-1$ and spin $s=t,
  t+1, \ldots$, and
\item the special FT fields with depth in the same range but spin
  $s=0,1,\ldots,t-1$.
\end{itemize}
We first sum the characters over the higher-depth FT field modules 
and obtain 
\begin{eqnarray}
  &&\sum_{t=1,3,\dots}^{2\ell-1} \sum_{s=t}^\infty
  \chi^{\phantom{g}}_{\cD(2;(s,s-t+1))}(q, \bm x) =
  \sum_{t=1,3,\dots}^{2\ell-1} \sum_{s=t}^\infty
  \chi^{\phantom{g}}_{\cD(1-s;(s-t))}(q, \bm x) + \nn && \qquad +
  \Big(\chi^{\phantom{g}}_{\rac_{\ell}}(q^{-1}, \bm x) \Big)^2 -
  \Big(\chi^{\phantom{g}}_{\rac_{\ell}}(q, \bm x) \Big)^2 \nn &&
  \qquad - \,\sum_{t=1,3,\dots}^{2\ell-1} \sum_{s=0}^{t-1} \left[
    \chi^{\phantom{g}}_{\cV(s-t+d-1;(s))}(q^{-1}, \bm x)-
    \chi^{\phantom{g}}_{\cV(s-t+d-1;(s))}(q,\bm x) \right].
  \label{A ell sum}
\end{eqnarray}
Again the second line vanishes due to the property
\begin{equation}
  \chi^{\phantom{g}}_{\rac_\ell}(q^{-1}, \bm x) = (-1)^{d+1}
  \chi^{\phantom{g}}_{\rac_\ell} (q, \bm x)\,,
\end{equation}
and in the third line we find the character
corresponding to the module of the special FT fields: 
\begin{eqnarray}
  \chi^{\phantom{g}}_{\cD(1+t-s;(s))}(q, \bm x) \eq
  \chi^{\phantom{g}}_{\cV(1+t-s;(s))}(q, \bm x) -
  \chi^{\phantom{g}}_{\cV(s-t+d-1;(s))}(q, \bm x) \nn \eq
  \chi^{\phantom{g}}_{\cV(s-t+d-1;(s))}(q^{-1}, \bm x) -
  \chi^{\phantom{g}}_{\cV(s-t+d-1;(s))}(q, \bm x)\,.
  \label{typeA s}
\end{eqnarray}
Here, we used again the property \eqref{inversion} for even
  $d$. The modules $\cD(1+t-s;(s))$ with $t=s,s+1,\ldots$ arise from
the ``non-standard'' BGG sequence of $\so(2,d)$, and they are
associated with the special FT fields.  This module exists in fact
only if $s \geqslant t-\frac d2+2$, which is the same condition that the
special FT field action has positive derivatives. This condition is
satisfied if $\ell<\frac d4$, and we obtain
\begin{eqnarray}
  &&\sum_{t=1,3,\dots}^{2\ell-1} \left[ \sum_{s=0}^{t-1}
    \chi^{\phantom{g}}_{\cD(1+t-s;(s))} + \sum_{s=t}^\infty
    \chi^{\phantom{g}}_{\cD(2;(s,s-t+1))} \right] =
  \sum_{t=1,3,\dots}^{2\ell-1} \sum_{s=t}^\infty
  \chi^{\phantom{g}}_{\cD(1-s;(s-t))}\,,
\end{eqnarray}
which confirms again the observation \eqref{main_identity} in the
type-A$_\ell$ cases with $\ell<\frac d4$.

Let us consider now the case $\ell\geqslant \frac d4$ where the third line
of \eqref{A ell sum} contains terms with $s \leqslant t-\frac d2+1$.
For $s=t-\frac d2+1$, the term simply vanishes. For $s \leqslant
t-\frac d2$, it becomes
\begin{equation}
  \chi^{\phantom{g}}_{\cV(s-t+d-1;(s))}(q^{-1}, \bm x) -
  \chi^{\phantom{g}}_{\cV(s-t+d-1;(s))}(q, \bm x) = -
  \chi^{\phantom{g}}_{\cD(s-t+d-1;(s))}(q,\bm x)\,,
\end{equation}
and cancels the characters $\chi^{\phantom{g}}_{\cD(1+t-s;(s))}$ with
$t-\frac d2 +2\leqslant s\leqslant \frac{t-d+1}2+\ell$ if
$\ell<\frac{d}2$.  In the end, for $\frac{d}4\leqslant \ell<\frac{d}2$
we find
\begin{equation}
  \sum_{t=1,3,\dots}^{2\ell-1} \left[ \sum_{s={\rm max}\{0,
      \frac{t-d+3}2+\ell\}}^{t-1} \chi^{\phantom{g}}_{\cD(1+t-s;(s))}
    + \sum_{s=t}^\infty \chi^{\phantom{g}}_{\cD(2;(s,s-t+1))}\right] =
  \sum_{t=1,3,\dots}^{2\ell-1} \sum_{s=t}^\infty
  \chi^{\phantom{g}}_{\cD(1-s;(s-t))}\,.
\end{equation}
The left-hand-side of the equality is precisely the linearized
on-shell spectrum of type-A$_\ell$ theory for $\frac{d}4\leqslant
\ell<\frac{d}2$\,,
and therefore confirms once again \eqref{main_identity}.

\subsubsection*{Going off-shell.}
The character of the off-shell depth-$t$ FT field with $1\leqslant t
\leqslant s$ is related to that of the on-shell one through
\begin{equation}
  \chi^{\phantom{g}}_{\Sh(1+t-s;(s))} =
  \chi^{\phantom{g}}_{\cD(2;(s,s-t+1))} +
  \chi^{\phantom{g}}_{\cD(s+d-t-1;(s))}\,,
  \label{gd}
\end{equation}
whereas that of the special FT field modules
$\cS(1+t-s;(s))=\cV(1+t-s;(s))$ with $t\geqslant s$ satisfy
\begin{equation}
  \chi^{\phantom{g}}_{\cS(1+t-s;(s))} =
  \chi^{\phantom{g}}_{\cD(1+t-s;(s))} +
  \chi^{\phantom{g}}_{\cV(s+d-t-1;(s))}\,.
  \label{ngd}
\end{equation}
Using \eqref{gd} and \eqref{ngd}, one can check that the relation
\eqref{twAdj_Adj_CHS} holds for any value of $\ell$, i.e. both of
$\ell < \frac{d}4$ and $\frac{d}4\le\ell< \frac d2$ cases.

\subsection{Character of type-B$_\ell$ theory}
\label{sec:typeBl}

Finally, let us consider the type-B$_\ell$ CHS theory based on the
order-$\ell$ spinor singleton which will be denoted Di$_\ell\,$. It is
related to the type-B$_\ell$ MHS theory in an analogous way to the
type-A$_\ell$ case.  Here, we focus on the case where the value of
$\ell$ is smaller enough than the dimension $d$ so that we do not
encounter any subtle issue analogous to what we found in the
type-A$_\ell$ theory.

The field content of the type-B$_\ell$ CHS theory,
\ba	\label{type B l}
	&& \textrm{Type-B$_\ell$ CHS}=
	\bigoplus_{t=-2(\ell-1)}^{2(\ell-1)}
	2\,\cD(1+t;\bm 0)\oplus\nn
	&&\oplus
	\bigoplus_{t=1}^{2\ell-1}
	2(2-\delta_{t,2\ell-1})
	\Bigg[
	\bigoplus_{s=1}^{t-1}
	\cD\big(1-s+t;(s)\big)
	\oplus
	\bigoplus_{s=t}^\infty
	\cD\big(2;(s,s-t+1)\big)\\
	&&\hspace{50pt}
	\oplus \bigoplus_{m=1}^{r-1}\bigg\{
	\bigoplus_{s=1}^{t}
	\cD\big(1-s+t;(s,\bm1^m)\big)
	\oplus
	\!\!\bigoplus_{s=t+1}^\infty\!\!
	\cD\big(1;(s,s-t+1,\bm1^{m-1})\big)\bigg\}
	\Bigg],\nonumber
\ea
can be read off from
that of the the type-B$_\ell$ MHS theory, which in turn is given by
the corresponding Flato-Fronsdal theorem \cite{Basile:2014wua} (see also \cite{Grigoriev:2018wrx} for the field-theoretical approach):
\begin{equation}
  {\chi^{\phantom{g}}_{\di_\ell}}^2= \sum_{t=-2(\ell-1)}^{2(\ell-1)}
 2\, \chi^{\phantom{g}}_{\cD(d-t-1;\zero)} + \sum_{t=1}^{2\ell-1}\,
  2(2-\delta_{t,2\ell-1})\, \sum_{m=0}^{r-1}\sum_{s=1}^\infty
  \chi^{\phantom{g}}_{\cD(s+d-t-1;(s,\1^m))}\,.
  \label{FF B}
\end{equation}
The universal factor 2 arises because we consider the parity-invariant
Di module which possesses the pseudo higher-spin currents with
$\gamma_{d+1}$ insertion.  Note that the character
$2\,\chi^{\phantom{g}}_{\cD(s+d-t-1;(s,\1^{r-1}))}$ should be
understood as $\chi^{\phantom{g}}_{\cD(s+d-t-1;(s,\1^{r-1}_+))} +
\chi^{\phantom{g}}_{\cD(s+d-t-1;(s,\1^{r-1}_-))}$\,.  This theory
contains infinitely many higher-spin gauge fields corresponding to the
modules $\cD(s+d-t-1;(s, \1^m))$ with $s\geqslant t+1-\delta_{m,0}$,
whose Killing tensors form the type-B$_\ell$ higher-spin algebra.
Note that for $s<t+1-\delta_{m,0}$, the generalized Verma module is
irreducible: $\cD(s+d-t-1;(s, \1^m))=\cV(s+d-t-1;(s, \1^m))$\,.  The
vector space of the type-B$_\ell$ higher-spin algebra can be
decomposed into $\so(2,d)$-modules as \cite{Vasiliev:2004cm}
\begin{equation}
  \textrm{Type-B}_\ell \ {\rm Algebra} \ \simeq
  \ \bigoplus_{t=1}^{2\ell-1} (2-\delta_{t,2\ell-1})\,
  \bigoplus_{m=0}^{r-1} \bigoplus_{s=t+1-\delta_{m,0}}^\infty
  \ \bracket[-24pt] {\footnotesize
    \gyoung(;_5{s-1},;_4{s-t},|{2}\vdts,;)}\,.
  \label{typeBl}
\end{equation}
The character of the conformal Killing tensors appearing above is
related to the character of the corresponding FT field module in M$_d$
through
\begin{eqnarray}
 && \chi^{\phantom{g}}_{\cD(1;(s,s-t+1,\1^{m-1}
  ))}(q, \bm x)
  =
   \chi^{\phantom{g}}_{\cD(1-s; (s-t,\1^m
  ))}(q, \bm x)+\nn &&\qquad\qquad +\,
  \chi^{\phantom{g}}_{\cD(s+d-t-1;(s,\1^m
  ))}(q^{-1}, \bm x)
  -
  \chi^{\phantom{g}}_{\cD(s+d-t-1;(s,\1^m
  ))}(q, \bm x) \,,
  \label{typeB id}
\end{eqnarray}
where $m\geqslant 1$ and $s\geqslant t+1$. If the latter condition is not
satisfied, then we get the special FT field module:
\begin{equation}
  \chi^{\phantom{g}}_{\cD(1-s+t;(s,\1^m))}(q, \bm x)=
  \chi^{\phantom{g}}_{\cV(s+d-t-1;(s,\1^m))}(q^{-1}, \bm
  x)-\chi^{\phantom{g}}_{\cV(s+d-t-1;(s,\1^m))}(q, \bm x)\,,
  \label{typeB s}
\end{equation}
We sum the characters \eqref{typeB id} and \eqref{typeB s} together
with the $m=0$ counterparts \eqref{decompo_CHS_PM} and \eqref{typeA s}
over the field content of type-B$_\ell$ CHS theory \eqref{type B l},
and use the Flato-Fronsdal theorem \eqref{FF B} to simplify the
series. In the end, we find the relations \eqref{main_identity} and
\eqref{twAdj_Adj_CHS} hold also for the type-B$_\ell$ theory.

\section{Discussion}
\label{sec:discu}

In this paper, we calculated the $\so(2,d)$ character of CHS gravity
spectrum and showed that it coincides with that of its global symmetry
algebra, namely (conformal) higher-spin algebra. The evaluation of the
full character requires the summation of the characters of each field
appearing in the spectrum of CHS gravity. The character of each
conformal gauge field satisfies the relation \eqref{rel_KT_FT_PM},
which provides a simple link to its associated conformal Killing
tensor, the collection of which span the CHS symmetry algebra. Our key
observation is that, in the character of each gauge field, the
contributions besides the Killing tensor cancel out after the
summation. Below, let us make two remarks: first about the key
relation \eqref{rel_KT_FT_PM}, then about the summation.

The relation \eqref{rel_KT_FT_PM} --- whose concrete versions for
partially-massless fields of symmetric and hook-symmetry types are
given respectively in \eqref{decompo_CHS_PM} and in \eqref{typeB id}
--- is in fact very general. In \hyperref[sec:typeBl]{Appendix
  \ref{app:zoo}}, we established the relation for partially-massless
fields of any depth and any symmetry: see \eqref{gen rel mix}. Note
that this relation is valid also for non-gauge fields --- e.g. for
totally-symmetric and hook-symmetry type fields we have \eqref{typeA
  s} and \eqref{typeB s} --- where only the Killing tensor
contribution is absent: see \eqref{sp FT ch} for the most general
case. Another way to state this property is that the character of any
$d$-dimensional conformal field, irrespective of whether it has gauge
symmetry or not, is given by the difference of the characters of the
associated field in AdS$_{d+1}$ with different boundary
conditions. The character for the field with Neumann condition is the
same as the character with Dirichlet condition but $q$ replaced by
$q^{-1}$, plus --- if the field has gauge symmetries --- the character
of the Killing tensor.  Therefore, the character of a theory composed
of multiple conformal fields is always given by the character of its
global symmetries and the rest which is the difference of the
character of the AdS$_{d+1}$ counterpart theory under the flip $q\to
q^{-1}$.  Hence, when the character of the AdS$_{d+1}$ theory is
symmetric under $q\to q^{-1}$, only the character of the global
symmetries survives as is the case for type-A$_\ell$ and
type-B$_\ell$. For this reason, our observation extends neither to
the type-C case nor to the minimal theories, but holds for the type-AB case \cite{Giombi:2016pvg} and type-AZ
theory \cite{Bae:2016xmv}.

Let us turn to the issue of the summation over spins or field
contents. Since we have just equated, as series, the character of CHS
gravity and that of its symmetries, we did not elaborate on whether
this series is convergent or not. In fact, there are several issues.
First, the character of the AdS$_{d+1}$ theory is given by a series
which is convergent inside the disk $|q|<1$. It shows the symmetry
$q\to q^{-1}$ only after an analytic continuation, and hence the
cancellation of these terms already involves a regularization.  Next,
the character of the higher-spin symmetries is given by a series which
does not have any region of convergence. Nevertheless, we can still
evaluate this series dividing it into the pieces having different
regions of convergence, and this procedure makes sense only in terms
of distributions (see \cite{Basile:2018dzi} for more discussions).
After these regularizations, we obtain the character of CHS gravity as
a function of $q$ and $\bm x$\,. For instance, for the type-A CHS
theory in $d=4$ we find
\begin{equation}
  \chi^{\so(2,4)}_{\rm\sst CHS}(q,x_1,x_2) = -\left(
  \frac{q\,(1-q^2)}{(1-x_1\,q)(1-x_1^{-1}\,q)(1-x_2\,q)(1-x_2^{-1}\,q)}
  \right)^2 +\,(q \leftrightarrow x_1) + (q \leftrightarrow x_2)\,.
  \label{chs ex}
\end{equation}
Since the one-loop partition function is the character evaluated at
$\bm x = \bm 1$, we can attempt to get the CHS partition function from
this character.  The limit $x_1, x_2\to1$ of the last two terms
of \eqref{chs ex} is singular, so we would need to subtract the finite
part in the limit.  However, this procedure is somewhat ambiguous as
it depends on how we take this limit.  Moreover, the simplest trials
(e.g. sending $x_1$ and $x_2$ consecutively or simultaneously to $1$)
do not easily reproduce the result found in \cite{Beccaria:2014jxa}:
\begin{equation}\label{partitionBBT}
  \cZ_{\rm\sst CHS}(q) = -\frac{q^2\,(11+26q+11q^2)}{6\,(1-q)^6}\,.
\end{equation}
The above was obtained by summing the partition functions of each FT
field (i.e. their individual characters evaluated in $\bm x = \bm 1$)
with a damping parameter $\epsilon$, i.e.
\begin{equation}
  \cZ_{\rm\sst CHS}(q) := \oint_{\mathscr C} \frac{\dd
    \epsilon}{2\pi\,i\,\epsilon}\, \sum_{s=0}^\infty
  e^{-\epsilon(s+\frac{d-3}2)}\,
  \chi^{\so(2,4)}_{\cD(2;(s,s)_0)}(q, \bm x) \big|_{\bm x = \bm
    1}\,.
\end{equation}
Here $\mathscr C$ is a contour encircling the origin of the complex
$\epsilon$-plane, so the contour integral picks up the finite part of
the integrand.  The shift $\frac{d-3}2$ in the
$e^{-\epsilon(s+\frac{d-3}2)}$ regularization is introduced to ensure
the vanishing of the $a$-anomaly coefficient and the Casimir energy of
CHS gravity \cite{Tseytlin:2013jya, Giombi:2014yra}. As already
detailed in \cite{Beccaria:2014jxa}, a closer investigation reveals
that the subtleties lie in the first term of the on-shell FT field
character,
 \begin{equation}
   \chi^{\so(2,4)}_{\cD(2;(s,s))}(q, x_1, x_2) =
   \chi^{\so(2,4)}_{\cV(2;(s,s)_0)}(q, x_1, x_2) -2\,
   \chi^{\so(2,4)}_{\cV(s+2;(s))}(q, x_1, x_2) +2\,
   \chi^{\so(2,4)}_{\cV(s+3;(s-1))}(q, x_1, x_2)\,,
\end{equation}
because the second and third terms in the above expression lead to
convergent series as we sum over the spins.  The first term is the
character of the generalized Verma module with lowest weight
$[2;(s,s)_0]$, so the spin-dependent part is simply the $\so(4)$
character,
\begin{equation}
  \chi^{\so(4)}_{(s,s)_0}(x_1,x_2) = \chi^{\so(3)}_{s}(x_1\,x_2) +
  \chi^{\so(3)}_{s}(x_1\,x_2^{-1})\,.
\end{equation}
Therefore, we see that the issue is in fact how to sum the above
character evaluated at $x_1=x_2=1$ over all spins. For a concrete
understanding, let us consider
\begin{equation}
  \sum_{s=0}^\infty \chi^{\so(3)}_{s}(x)\, e^{-\e\,(s+\frac12)} =
  \frac{e^{-\frac12\,\e}\,(1+e^{-\e})\,x}{(e^{-\e}-x)(-1+e^{-\e}\,x)}\,.
\end{equation}
The above function is regular in either limit $x\to 1$ or $\e\to 0$
but not the simultaneous limit. And depending on the evaluation order,
we indeed obtain different results. Setting $x$ to $1$ and then
extracting the finite part in the $\epsilon \rightarrow 0$ expansion
leads to the partition function \eqref{partitionBBT}, whereas setting
first $\epsilon$ to $0$ and then extracting the finite part in the
$x\rightarrow1$ expansion leads to
\begin{equation}
  \cZ_{\rm\sst CHS}(q) = - \frac{2\, q^2\,(1+4q+q^2)}{(1-q)^6}\,.
\end{equation}
It is interesting to note that the above partition function also gives
vanishing Casimir energy. The cancellation of the $a$-anomaly of CHS
gravity can be also shown using the characters (see
\hyperref[char_on-shell]{Subsection \ref{char_on-shell}}).

Let us conclude by a few comments on CHS theories around an AdS
background.  The kinetic operator of spin-$s$ Fradkin-Tseytlin fields
around AdS$_d$ factorize into a product of kinetic operators of
spin-$s$ partially-massless fields with all possible depths (together
with a finite number of spin-$s$ massive fields in dimensions
$d\geqslant4$) \cite{Metsaev:2014iwa, Nutma:2014pua,
  Joung:2012qy}. Notice that this property was recently generalized to
the case of higher-depth Fradkin-Tseytlin fields in
\cite{Grigoriev:2018mkp}. From a group-theoretical perspective, the
factorization of the kinetic operator of Fradkin-Tseytlin fields is
related to the branching rule of the corresponding module: as detailed
in \hyperref[app:branching]{Appendix \ref{app:branching}}, the
on-shell Fradkin-Tseytlin module branches into the $\so(2,d-1)$
modules which correspond to the kernels of each factor of the
Fradkin-Tseytlin kinetic operator.\footnote{A noteworthy subtlety here
  is that here appear both of the modules corresponding to the
  solutions with Dirichlet and Neumann boundary conditions.}
Consequently, the CHS gravity around an AdS background can be viewed
as a \emph{local} interacting theory of massless, partially-massless
and massive (with ``the mass squared'' smaller than that of massless
fields) fields of arbitrary spins.  Such a field content as an AdS$_d$
theory can be obtained by simply branching the representation carried
by the CHS gravity spectrum from $\so(2,d)$ onto $\so(2,d-1)$. For
instance, in $d=4$ dimensions the type-A CHS theory spectrum branches
onto the direct sum of partially-massless fields in AdS$_4$ with all
integer spins and depths.\footnote{When $d\geqslant6$, the branching
  rule will also contain an infinite number of massive fields with
  arbitrarily high integer spin.}  Taking a definite boundary
condition\footnote{When $d=4$, the partially-massless fields with
  Dirichlet and Neumann boundary conditions correspond to isomorphic
  modules.} we find that the $\so(2,3)$ character of this theory reads
\begin{equation}
  \sum_{s=1}^\infty \sum_{t=1}^s
  \chi^{\so(2,3)}_{\D(s+2-t;(s))}(\beta,\alpha) =
  \frac{\sinh^2\frac\beta2 \cos\alpha -
    \sin^2\frac\alpha2}{8\,(\cosh\beta - \cos\alpha)^2\,
    \sin^2\frac\alpha2\, \sinh^2\frac\beta2}\,,
  \label{char_derived_CHS}
\end{equation}
where we have written the character in terms of the temperature
$\beta$ and the variable $\alpha$ defined through $q = e^{-\beta}$ and
$x=e^{i\alpha}$.  Using the method introduced in
\cite{Bae:2016rgm,Basile:2018zoy}, we can compute the one-loop free
energy of this theory in Euclidean AdS$_4$ using the above
character. To do so, one needs to evaluate the character
\eqref{char_derived_CHS} and its derivatives (with respect to the
variable $\alpha$) in $\alpha\rightarrow0$, which is singular here
(contrary to MHS theories and their partially-massless cousins). As a
consequence, one has to introduce a new regularization scheme to cure
this additional source of divergence.  We can easily think of two
possibilities:
\begin{itemize}
\item To extract only the finite part of \eqref{char_derived_CHS} in
  its $\alpha\rightarrow0$ expansion which leads to the one-loop free
  energy
  \begin{equation}
    \Gamma^{\sst (1)}_{\sst\text{AdS}_4} = - \frac{\zeta(3)}{192\pi^2}
    - \frac{\zeta(5)}{48\,\pi^4} - \frac{\zeta(7)}{64\,\pi^6}\,.
  \end{equation}
\item To introduce a damping factor in the summation of the character,
  \begin{equation}
    \sum_{s=1}^\infty \sum_{t=1}^s\, e^{-\e(s+\gamma)}\,
    \chi^{\so(2,3)}_{\cD(s+2-t;(s))}(\beta,\alpha)\,,
    \label{result 1}
  \end{equation}
  so that the resulting character has a finite limit in
  $\alpha\to0$. One can then extract the finite part in the
  $\e\rightarrow0$ expansion and compute the one-loop free energy
  using the resulting expression. Using a shift $\gamma=\frac12$ as in
  the case of MHS or CHS theory in four dimensions, we end up with
  \begin{equation}
    \Gamma^{\sst (1)}_{\sst\text{AdS}_4} = -
    \frac{\zeta(3)}{7680\,\pi^2} - \frac{\zeta(5)}{384\,\pi^4}\,.
    \label{result 2}
  \end{equation}
\end{itemize}
None of the results \eqref{result 1} and \eqref{result 2} is
particularly convincing, and hence we would need clearer guidelines
for the regularization from other inspections.  This naturally calls
for a better understanding of CHS gravity from the point of view of an
``all-depth'' partially-massless higher-spin theory in AdS$_d$.  In
fact the latter perspective is interesting to explore in its own right
and there are several amusing questions.  The first question is
whether CHS gravity can afford a truncation like the one from Weyl
gravity to Einstein gravity \cite{Maldacena:2011mk,
  Anastasiou:2016jix}. The truncation to MHS fields would not be
consistent since there is no subalgebra of the CHS algebra compatible
with the truncated spectrum, but the truncation to all odd-depth
fields may lead to a consistent theory \cite{Joung:2015jza}.  The
second question is what might be the CFT$_{d-1}$ dual of CHS gravity
viewed as an exotic partially-massless higher-spin theory in AdS$_d$.
It cannot be one of the usual free CFTs as they do not exhibit such
operator spectra, but their exotic cousins might be decent candidates:
in \cite{Joung:2015jza} the non-local free scalar CFT with the kinetic
operator $\sqrt{\Box}$ was proposed as the dual of the odd-depth
truncation of CHS gravity in AdS.  The third question is whether CHS
gravity can be defined as a local theory in certain odd dimensions. It
is well known that three-dimensional CHS gravity can be realized as a
Chern-Simons theory.  Therefore, it is natural to inquire whether the
higher-dimensional Chern-Simons actions may prove useful for the local
realization of some parity-breaking CHS gravity in odd dimensions.
Lastly, assuming this works, one might then speculate whether one can
iterate the construction \cite{Beccaria:2016tqy}: MHS gravity in
AdS$_{d+1}$ leads to CHS gravity in AdS$_d$, which in turn could lead
to another class of CHS gravity in AdS$_{d-1}$, etc.

\acknowledgments

X.B. thanks M.~Grigoriev for illuminating discussions on the field-
versus group-theoretical description of shadow fields. X.B. is also
grateful to the organizers of the conference ``Gauge/Gravity Duality
2018'' (W\" urzburg, 30 July - 03 August) where some of the present
results were announced. The research of T.B. and E.J. was supported by
the National Research Foundation (Korea) through the grant
2014R1A6A3A04056670. The research of T.B. was supported by Korea
Research Fellowship Program through the National Research Foundation
of Korea (NRF) funded by the Ministry of Science and ICT
2018H1D3A1A02074698. The research of X.B. was supported by the Russian
Science Foundation grant 14-42-00047 in association with the Lebedev
Physical Institute.

\appendix
\section{Zoo of modules}
\label{app:zoo}
In this appendix, we review the description of the various fields of
interest for this paper in terms of the corresponding
$\so(2,d)$-modules, as well as the associated characters. To do so
systematically, we go through the BGG resolution of these modules,
following the discussions in \cite{Lepowsky1977, Shaynkman:2004vu,
  Beccaria:2014jxa} (see also \cite{Dolan:2005wy, Bourget:2017kik} for
more details on the relevant characters from a different perspective).

\begin{itemize}
\item {\bf AdS$_{d+1}$ gauge field module / CFT$_d$ current:} The
  module defined by the lowest weight
  \begin{equation}
    [s_{p}+d-p-t\,;\, (s_1, \dots, s_{p}, s_{p+1}, \dots,
      s_r)]\,\, , \quad 1 \leqslant t \leqslant s_{p} - s_{p+1}\,,
  \end{equation}
  corresponds to a partially-massless
  mixed-symmetry field with spin $\Y = (s_1, \dots, s_r)$ and minimal
  energy $\Delta_{p}^{(t)} := s_{p}+d-p-t$. This irreducible module is
  defined as the quotient of the generalized Verma module induced by
  the $\so(2)\oplus\so(d)$ module of weight $[\Delta^{(t)}_{p}\,;\,
    \Y]$, by its maximal submodule. The latter is the module whose
  lowest weight reads
  \begin{equation}
    [s_{p}+d-p; (s_1, \dots, s_{p-1}, \pos{s_{p}-t}{p\th},
      s_{p+1}, \dots, s_r)]\, .
  \end{equation}
  The presence of this submodule signals the fact that these partially
  massless fields are subject to gauge symmetries, namely the
  submodule of lowest weight $[\Delta^{(t)}_{p} + t\,;\,
    \Y_{p,t}]$ with
  \begin{equation}
    \Y_{p,t} := (s_1, \dots, s_{p-1}, s_{p}-t, s_{p+1}, \dots,
    s_r)\, .
  \end{equation}
  The resulting quotient,
  \begin{equation}
    \D\big( \Delta^{(t)}_{p}\,;\, \Y\big) = \frac{\V\big(
      \Delta^{(t)}_{p}\,;\, \Y\big)}{\D\big( \Delta^{(t)}_{p} +
      t\,;\, \Y_{p,t}\big)}\, ,
  \end{equation}
  can be realized as the space of traceless and divergenceless tensor
  with symmetry $\Y$ and solution to the wave equation
  \begin{equation}
    \big( \nabla^2 - m^2_{p,t} \big)\, \varphi_\Y = 0\, , \qquad
    m^2_{p,t} := \Delta_{p}^{(t)}(\Delta_{p}^{(t)}-d) -
    \sum_{k=1}^r s_k\, , 
  \end{equation}
  modulo the gauge transformations
  \begin{equation}
    \delta_\xi \varphi_\Y = \nabla^t\, \xi_{\Y_{p,t}}\, ,
    \label{gauge}
  \end{equation}
  where $\xi_{\Y_{p,t}}$ is also a traceless and divergenceless
  tensor with symmetry $\Y_{p,t}$ and subject to a wave equation
  similar that of $\varphi_\Y$ (see \cite{Metsaev:1995re,
    Metsaev:1997nj, Boulanger:2008up, Boulanger:2008kw,
    Skvortsov:2009zu, Skvortsov:2009nv} for more details). When $p >
  1$, the gauge symmetry \eqref{gauge} is reducible, i.e. it is
  trivial for some particular class of gauge parameters. This implies
  that the module $\cD\big(\Delta^{(t)}_{p}+t\,;\, \Y_{p,t}\big)$ of gauge parameters is itself a quotient, obtained
  by modding out of $\cV\big(\Delta^{(t)}_{p}+t\,;\, \Y_{p,t}\big)$
  the irreducible submodule describing the parameters leading to
  trivial gauge transformations. For a generic partially-massless as
  we are considering here, there are $p$ classes of ``gauge
  parameters'', the genuine one with diagram $\Y_{p,t}$ and
  $(p-1)$ reducibilities, which are obtained from the BGG
  resolution. Denoting $\Y^{(k)}_{p,t}$ the Young diagram describing the $k$th
  of these reducibility parameters, they can be expressed in terms of
  $\Y_{p,t}$ by
  \begin{equation}
    \Y_{p,t}^{(k)} = (s_1, \dots, s_{p-k-1},
    \pos{s_{p-k}-1-n_{k,p}}{(p-k)\th}, \dots,
    s_{p-1}-1-n_{1,p}, \pos{s_{p}-t}{p\th}, s_{p+1}, \dots,
    s_r)
  \end{equation}
  with
  \begin{equation}
    n_{j,p} := s_{p-j}-s_{p-j+1}\,, \quad 1 \leqslant k
    \leqslant p-1
  \end{equation}
  with by convention $n_{0,p}=0$. Defining
  \begin{equation}
    \nu_{k,p} := \sum_{j=1}^k n_{k,p}\,, \quad 1 \leqslant k
    \leqslant p-1\,,
  \end{equation}
  with again the convention that $\nu_{0,p}=0$, the minimal energy
  of these reducibility parameters reads
  \begin{equation}
    \Delta_{p}^{(t)} + t + k + \nu_{k,p}\,.
  \end{equation}
  Schematically, each reducibility parameter is obtained from the
  previous one by removing more and more boxes. Starting from the
  gauge parameter, where $t$ boxes are removed from the $p$th row,
  one then obtain the first reducibility parameter by removing boxes
  from the row above (the ($p-1$)th one) until this row has one less
  box than the one below (the $p$th one) in the original Young
  diagram (i.e. $s_{p}-1$). Applying the same procedure to the
  obtained reducibility parameter, one can construct the next one and
  so on.

  From the $d$-dimensional point of view, these modules describe
  (partially-)conserved currents of spin $\Y$, i.e. operators of
  conformal weight $\Delta_{p}^{(t)}$ which are traceless tensors
  with the symmetry properties of the Young diagram $\Y$. The
  conservation law obeyed by these operators is of the form
  \begin{equation}\label{conservationlaw}
    \proj^{\Y_{p,t}}_{\rm \sst T}\big( \partial^t J_\Y \big) \approx
    0\,,
  \end{equation}
  where $\proj^{\Y_{p,t}}_{\rm \sst T}$ is a traceless projector
  onto $\Y_{p,t}$.  In this context the submodules involved in the
  definition of this irreducible module are interpreted slightly
  differently than in the AdS$_{d+1}$ case. The first submodule is
    spanned by the conservation law \eqref{conservationlaw}, i.e. a
    suitable projection of a (or several) divergence(s), and all its
    descendants. The remaining submodules correspond to descendants of
    the conservation law which are identically\,\footnote{For
        instance, the module $\D(d-3;(1,1)$) corresponds to an
        antisymmetric tensor $F_{ab}=-F_{ba}$ obeying the conservation law
        $\partial^b\, F_{ab} \approx 0$. Its first submodule is
        $\D(d-2;(1))$ correspond to the image of the divergence of
        $F_{ab}$ in $\D(d-3;(1,1))$, while the second submodule
        $\D(d-1;(0))$ is the image of the trivially conserved quantity
        $\partial^a\partial^b\, F_{ab}$ in $\D(d-2;(1))$.} vanishing
    (i.e. for symmetry reasons).

  We can now write down (irrespectively of the parity of $d$) the
  character of the module gauge field module $\D\big(
  \Delta^{(t)}_{p}\,;\, \Y\big)$ as follows:
  \begin{equation}
    \chi^{\so(2,d)}_{\D(\Delta^{(t)}_{p}; \Y)}(q, \bm x) = \Big(
    q^{\Delta_{p}^{(t)}}\, \chi^{\so(d)}_{\Y}(\bm x)\, +
    \sum_{k=0}^{p-1} (-1)^{k+1}\,
    q^{\Delta_{p}^{(t)}+t+k+\nu_{k,p}}\,
    \chi^{\so(d)}_{\Y_{p,t}^{(k)}}(\bm x) \Big) \Pd d (q, \bm x)\,,
    \label{character_PM}
  \end{equation}
  with the convention that $\Y^{(0)}_{p,t} = \Y_{p,t}$.
  
\item {\bf Curvature of the gauge field in AdS$_{d+1}$:} The module
  whose lowest weight reads
  \begin{equation}
    [s_{p+1}+d-p-1; (s_1, \dots, s_{p},
      \pos{s_{p}-t+1}{(p+1)\th}, s_{p+2}, \dots, s_r)]
  \end{equation}
  corresponds to the module of the curvature $R_{\Y_c}$ of the gauge
  field $\varphi_\Y$. It is obtained by taking $s_{p}-s_{p+1}-t+1$
  derivative of $\varphi_\Y$ and projecting them so that the resulting
  object as the symmetry property of the Young diagram $\Y_c$ which is
  obtained from $\Y$ by adding $s_{p}-s_{p+1}-t+1$ to the
  $(p+1)$th row
  \begin{equation}
    \Y_c = (s_1,\dots, s_{p}, \pos{s_{p}-t+1}{(p+1)\th},
    s_{p+2}, \dots, s_r)\, .
  \end{equation}
  Schematically, the curvature is given by
  \begin{equation}
    R_{\Y_c} = \proj^{\Y_c} \Big(
    \partial^{\Delta_{p}^{(t)}-\Delta_c} \phi_\Y \Big)\,,
  \end{equation}
  where $\proj^{\Y_c}_{\rm\sst T}$ is the projector onto the symmetry
  of the Young diagram $\Y_c$ (without any tracelessness condition).
  Accordingly, the minimal energy of the module of the curvature
  tensor is given by
  \begin{equation}
    \Delta_c  = s_{p+1} + d - p - 1 \equiv
    \Delta^{(t)}_{p} + s_{p+1}-s_{p}+t-1\, ,
  \end{equation}
  i.e. the minimal energy of the gauge field $\varphi_\Y$ added by the number of derivative acting on it to make up the
  curvature. The character of this module reads (irrespectively of the
  parity of $d$)
  \begin{eqnarray}
    \chi^{\so(2,d)}_{\D(\Delta_c; \Y_c)}(q, \bm x) & = & \Big(
    q^{\Delta_c} \chi^{\so(d)}_{\Y_c}(\bm x) -
    q^{\Delta_{p}^{(t)}}\, \chi^{\so(d)}_{\Y}(\bm x)\, \\ && \qquad
    \qquad \qquad + \sum_{k=0}^{p-1} (-1)^{k}\,
    q^{\Delta_{p}^{(t)}+t+k+\nu_{k,p}}\,
    \chi^{\so(d)}_{\Y_{p,t}^{(k)}}(\bm x) \Big) \Pd d (q, \bm
    x)\,. \nonumber
  \end{eqnarray}
  
\item {\bf Conformal Killing tensor:} The finite-dimensional module
  whose BGG resolution contains the above gauge field module
  corresponds to that of its conformal Killing tensor. In AdS$_{d+1}$,
  this corresponds to the space of solution of
  \begin{equation}
    \proj^\Y \Big( \nabla^t \xi_{\Y_{p,t}} \Big) = 0\,,
  \end{equation} 
  where $\proj^{\Y}$ projects onto the symmetry of the Young diagram
  $\Y$ without any tracelessness condition, whereas in M$_d$ this
  module corresponds to the space of solution of the conformal Killing
  equation
  \begin{equation}
    \proj_{\rm\sst T}^\Y \Big( \partial^t \xi_{\Y_{p,t}} \Big) =
    0\,.
  \end{equation}
  In terms of $\so(2,d)$-weight, the conformal Killing tensor module
  has lowest weight
  \begin{equation}
    [1-s_1\,;\, (s_2-1, \dots, s_{p}-1, s_{p}-t, s_{p+1}, \dots,
      s_{r-1}, (-)^{d+1}s_r)]\,.
  \end{equation}
  Written in terms of its $\so(2)\oplus\so(d)$ components, this
  character reads
  \begin{eqnarray}
    && \chi^{\so(2,d)}_{\D(1-s_1; (s_2-1, \dots, s_{p}-1, s_{p}-t,
        s_{p+1}, \dots, s_{r-1}, (-)^{d+1} s_r))}(q, \bm x)
    = \label{character_Killing} \\ && \qquad \qquad \Pd d (q, \bm
    x)\,\Bigg[ \sum_{k=1}^{p} (-1)^{k-1}\, \Big(
      q^{d-\Delta_{p}^{(t)}-t-p+k-\nu_{p-k,p}}\,
      \chi^{\so(d)}_{\Y_{p,t,-}^{(p-k)}}(\bm x) \nn
      && \hspace{200pt} + (-)^d
      q^{\Delta_{p}^{(t)}+t+p-k+\nu_{p-k,p}}\,
      \chi^{\so(d)}_{\Y_{p,t,+}^{(p-k)}}(\bm x) \Big) \nn &&
      \qquad \qquad \qquad + (-1)^{p} \Big(
      q^{d-\Delta_{p}^{(t)}}\, \chi^{\so(d)}_{\Y_-}(\bm x) + (-)^d
      q^{\Delta_{p}^{(t)}}\, \chi^{\so(d)}_{\Y_+}(\bm x) \Big)\, \nn
      && + \sum_{m=0}^{r-p-1} (-1)^{m+p+1} \Big(
      q^{d-\Delta_c+\bar \nu_{m,p}}\,
      \chi^{\so(d)}_{\Y_{c,-}^{(m)}}(\bm x) + (-)^d q^{\Delta_c-\bar
        \nu_{m,p}}\, \chi^{\so(d)}_{\Y_{c,+}^{(m)}}(\bm x) \Big)
      \Bigg]\,,\nonumber
  \end{eqnarray}
  where
  \begin{equation}
    \Y_{c,\pm}^{(m)} = (s_1, \dots, s_{p},
    \pos{s_{p}-t+1}{(p+1)\th}, s_{p+2}+\bar n_{1,p}, \dots,
    \pos{s_{p+m+1}+\bar n_{m,p}}{(p+m+1)\th}, s_{p+m+2},
    \dots, \pm s_r)
  \end{equation}
  with
  \begin{equation}
    \bar n_{j,p} := s_{p+j} - s_{p+j+1}+1\, , \quad \bar
    \nu_{m,p} := \sum_{j=1}^m \bar n_{j,p}\, , \quad 1 \leqslant j
    \leqslant r-p-1\,,
  \end{equation}
  and by convention $\bar \nu_{0,p}=0$. Its character coincides with
  that of the finite-dimensional $\so(2+d)$ representation
  \begin{equation}
    (s_1-1, s_2-1, \dots, s_{p}-1, s_{p}-t, s_{p+1}, \dots,
    s_{r-1}, (-)^{d+1} s_r)\,,
  \end{equation}
  evaluated in $(q^{-1}, \bm x)$, i.e.
  \begin{eqnarray}
    \chi^{\so(2,d)}_{\D(1-s_1;(s_2-1,\dots, s_{p}-1, s_{p}-t,
        s_{p+1}, \dots, (-)^{d+1}s_r))}(q, \bm x) \qquad \qquad
    \qquad \qquad \\ =\, \chi^{\so(d+2)}_{(s_1-1,s_2-1,\dots,
      s_{p}-1, s_{p}-t, s_{p+1}, \dots,
      (-)^{d+1}s_r)}(q^{-1},\bm x)\,. \nonumber
  \end{eqnarray}
    
\item {\bf On-shell Fradkin-Tseytlin field in $M_d$\,:}
  The module defined by the lowest weight
  \begin{equation}
    [p+1-s_{p+1}; (s_1, \dots, s_{p},
      \pos{s_{p}-t+1}{(p+1)\th}, s_{p+2}, \dots, (-)^{d+1}s_r)]
  \end{equation}
  corresponds to the on-shell Weyl tensor $C_{\Y_c}$ of the
  Fradkin-Tseytlin module, which is obtained by acting with
  $\Delta_{p}^{(t)}-\Delta_c = s_{p}-s_{p+1}-t+1$ derivatives on
  $\phi_\Y$, projected so that the resulting tensor has the symmetry
  of $\Y_c$. Schematically, we have
  \begin{equation}
    C_{\Y_c} = \proj^{\Y_c}_{\rm\sst T} \Big(
    \partial^{\Delta_{p}^{(t)}-\Delta_c} \phi_\Y \Big)\,,
  \end{equation}
  where $\proj^{\Y_c}_{\rm\sst T}$ is the traceless projector onto the
  symmetry of the Young diagram $\Y_c$.  The character of this module
  reads, for $d=2r$:
  \begin{eqnarray}
    && \chi^{\so(2,d)}_{\D(d-\Delta_c;\Y_c)}(q, \bm x) = \\ && \qquad
    \Pd d (q, \bm x)\,\Bigg[ \sum_{m=0}^{r-p-1} (-1)^m \Big(
      q^{d-\Delta_c+\bar \nu_{m,p}}\,
      \chi^{\so(d)}_{\Y_{c,-}^{(m)}}(\bm x) + q^{\Delta_c-\bar
        \nu_{m,p}}\, \chi^{\so(d)}_{\Y_{c,+}^{(m)}}(\bm x)
      \Big) \label{character_Weyl} \nn && \qquad \qquad \qquad \qquad
      - 2\, \Big( q^{\Delta_{p}^{(t)}}\, \chi^{\so(d)}_{\Y}(\bm x)\,
      + \sum_{k=0}^{p-1} (-1)^{k+1}\,
      q^{\Delta_{p}^{(t)}+t+k+\nu_{k,p}}\,
      \chi^{\so(d)}_{\Y_{p,t}^{(k)}}(\bm x) \Big) \Bigg]\,,
    \nonumber
  \end{eqnarray}
  and for $d=2r+1$:
  \begin{eqnarray}
    && \chi^{\so(2,d)}_{\D(-\Delta_c;\Y_c)}(q, \bm x)
    = \label{character_Weyl} \\ && \qquad \qquad \Pd d (q, \bm x)\,
    \Bigg[ \sum_{m=0}^{r-p-1} (-1)^m \Big( q^{d-\Delta_c+\bar
        \nu_{m,p}}\, \chi^{\so(d)}_{\Y_{c,-}^{(m)}}(\bm x) -
      q^{\Delta_c-\bar \nu_{m,p}}\,
      \chi^{\so(d)}_{\Y_{c,+}^{(m)}}(\bm x) \Big) \nn && \qquad \qquad
      \qquad \qquad + q^{\Delta_{p}^{(t)}}\, \chi^{\so(d)}_{\Y}(\bm
      x)\, + \sum_{k=0}^{p-1} (-1)^{k+1}\,
      q^{\Delta_{p}^{(t)}+t+k+\nu_{k,p}}\,
      \chi^{\so(d)}_{\Y_{p,t}^{(k)}}(\bm x) \Bigg]\,. \nonumber
  \end{eqnarray}
  Notice that in both cases, characters of modules labelled by the
  $\so(d)$-weights $\Y_c^{(m)}$ correspond to the generalized
  Bianchi identities verified by the Weyl tensor.
  
\item {\bf Off-shell Fradkin-Tseytlin field in M$_d$\,:}
  The off-shell Fradkin-Tseytlin, or shadow, field $\phi_\Y$
  associated to the (partially) massless field $\varphi_\Y$ is a field
  of conformal weight
  \begin{equation}
    \Delta_{\phi_\Y} = d-\Delta_{p}^{(t)} = p + t - s_{p}\,,
  \end{equation}
  and spin $\Y$, subject to the gauge transformation
  \begin{equation}
    \delta_{\xi_, \sigma} = \proj^\Y \Big( \partial^t\,
    \xi_{\Y_{p,t}} + \eta\, \sigma_{\check \Y} \Big)\,,
    \label{gauge_shadow_mixed}
  \end{equation}
  where $\proj^\Y$ project onto the symmetry of $\Y$ whereas
  $\check\Y$ denotes any Young diagram obtained by taking a trace
    of $\Y$, so that $\sigma_{\check\Y}$ should be understood as a
    collection of Weyl gauge parameters with the symmetry of all
    possible traces of $\Y$. As already mentioned previously, this
  field does not correspond to a generalized Verma module and
  consequently is not found in the BGG resolution in even
  dimension. However, as proposed in \cite{Beccaria:2014jxa}, the corresponding
  character can still be computed by translating the
  field-theoretical definition of this conformal gauge field. In even $d$ dimensions,
  we therefore define
  \begin{equation}
    \chi_{\Sh(d-\Delta^{(t)}_{p}; \Y)}^{\so(2,d)}(q, \bm x) :=
    q^{d-\Delta_{p}^{(t)}}\, \chi^{\so(d)}_{\Y}(\bm x)\, \Pd d (q,
    \bm x) - \chi_{\U(d-\Delta^{(t)}_{p}; \Y_-)}^{\so(2,d)}(q, \bm
    x)\,,
  \end{equation}
  where:
  \begin{itemize}
  \item The first term, which is the character of the generalized
    module with lowest weight $(d-\Delta_{p}^{(t)};\Y)$ is meant to
    represent the fact that the shadow field is a tensor of symmetry
    $\Y$ and with conformal weight $d-\Delta_{p}^{(t)}$;
  \item The second term is the character of the module
    $\U(d-\Delta_{p}^{(t)};\Y_-)$. This module appears in the BGG
    resolution as the maximal submodule of
    $\V(d-\Delta_{p}^{(t)}-t\,;\, \Y_{p,t, -})$, and corresponds
    to the pure gauge modes of the shadow field associated to the
    gauge symmetry \eqref{gauge_shadow_mixed}.
  \end{itemize}
  The character $\chi_{\U(d-\Delta^{(t)}_{p}; \Y_-)}^{\so(2,d)}(q, \bm
  x)$ can be computed following the algorithm spelled out in
  \cite{Beccaria:2014jxa}, which yields
  \begin{eqnarray}
    \chi_{\U(d-\Delta^{(t)}_{p}; \Y_-)}^{\so(2,d)}(q, \bm x) & = & \Pd
    d (q, \bm x)\,\Bigg[ q^{d-\Delta_{p}^{(t)}}\,
      \chi^{\so(d)}_{\Y}(\bm x) \label{char_U} \\ && -
      \sum_{m=0}^{r-p-1} (-1)^m \Big( q^{d-\Delta_c+\bar \nu_{m,p}}\,
      \chi^{\so(d)}_{\Y_{c,-}^{(m)}}(\bm x) + q^{\Delta_c-\bar
        \nu_{m,p}}\, \chi^{\so(d)}_{\Y_{c,+}^{(m)}}(\bm x) \Big)
      \nonumber \\ && \qquad + q^{\Delta_{p}^{(t)}}\,
      \chi^{\so(d)}_{\Y}(\bm x)\, + \sum_{k=0}^{p-1} (-1)^{k+1}\,
      q^{\Delta_{p}^{(t)}+t+k+\nu_{k,p}}\,
      \chi^{\so(d)}_{\Y_{p,t}^{(k)}}(\bm x) \Bigg]\,. \nonumber
  \end{eqnarray}
  Upon using \eqref{char_U}, we can now write explicitely the
  character of the shadow field $\phi_\Y$ as
  \begin{eqnarray}
    && \chi_{\Sh(d-\Delta^{(t)}_{p}; \Y)}^{\so(2,d)}(q, \bm x)
    = \label{character_Sh} \\ && \qquad \qquad \Pd d (q, \bm
    x)\,\Bigg[ \sum_{m=0}^{r-p-1} (-1)^m \Big( q^{d-\Delta_c+\bar
        \nu_{m,p}}\, \chi^{\so(d)}_{\Y_{c,-}^{(m)}}(\bm x) +
      q^{\Delta_c-\bar \nu_{m,p}}\,
      \chi^{\so(d)}_{\Y_{c,+}^{(m)}}(\bm x) \Big) \nonumber \\ &&
      \qquad \qquad \qquad - \Big( q^{\Delta_{p}^{(t)}}\,
      \chi^{\so(d)}_{\Y}(\bm x)\, + \sum_{k=0}^{p-1} (-1)^{k+1}\,
      q^{\Delta_{p}^{(t)}+t+k+\nu_{k,p}}\,
      \chi^{\so(d)}_{\Y_{p,t}^{(k)}}(\bm x) \Big) \Bigg]\,.\nonumber
  \end{eqnarray}

  As pointed out in \hyperref[sec:typeA]{Section \ref{sec:typeA}}, the
  shadow field module in odd dimension is isomorphic to that of the
  off-shell Weyl tensor in odd dimensions, and hence their character
  are identical. It is worth noticing though that the $d=2r+1$ case
  can be describe in a similar way to the $d=2r$ case, by defining
  \begin{equation}
    \chi_{\Sh(d-\Delta^{(t)}_{p}; \Y)}^{\so(2,d)}(q, \bm x) :=
    q^{d-\Delta_{p}^{(t)}}\, \chi^{\so(d)}_{\Y}(\bm x)\, \Pd d (q,
    \bm x) - \chi_{\D(d-\Delta^{(t)}_{p}; \Y)}^{\so(2,d)}(q, \bm x)\,.
    \label{def_shadow_odd}
  \end{equation}
  The two terms here have the same interpretation as in the $d=2r$
  case, except the module of the pure gauge modes of the shadow field
  $\D(d-\Delta^{(t)}_{p};\Y)$ is irreducible in $d=2r+1$. As a
  consequence, this module is now part of the BGG resolution in the sense that it is
  isomorphic to the quotient
  \begin{equation}
    \D(d-\Delta_{p}^{(t)};\Y) \cong
    \frac{\V(d-\Delta_{p}^{(t)};\Y)}{\D(d-\Delta_c;\Y_c)}\,,
  \end{equation}
  where one can recognize the module corresponding to the off-shell Weyl tensor in the denominator
  of the above equation. Its character therefore reads
  \begin{eqnarray}
    \chi_{\D(d-\Delta^{(t)}_{p}; \Y)}^{\so(2,d)}(q, \bm x) =
    q^{d-\Delta_{p}^{(t)}}\, \chi^{\so(d)}_{\Y}(\bm x)\, \Pd d (q,
    \bm x) - \chi^{\so(2,d)}_{\D(d-\Delta_c;\Y_c)}(q, \bm
    x)\,,
  \end{eqnarray}
  and hence using the definition \eqref{def_shadow_odd} we recover the
  expected result
  \begin{eqnarray}
    \chi_{\Sh(d-\Delta^{(t)}_{p}; \Y)}^{\so(2,d)}(q, \bm x) =
    \chi^{\so(2,d)}_{\D(d-\Delta_c;\Y_c)}(q, \bm x)\,,
  \end{eqnarray}
    in accordance with the fact that the module of the shadow field is
    isomorphic to that of the off-shell Weyl tensor.
\end{itemize}

\paragraph{Series of modules with non-integral weight.}
Another class of irreducible modules, with non-integral weight, can
exists and are defined by the quotient
\begin{equation}
  \D\big( \Delta\,;\, \Y\big) \cong \frac{\V\big( \Delta\,;\,
    \Y_+\big)}{\D\big( d-\Delta\,;\, \Y_-\big)}
  \label{nonintegral_module}
\end{equation}
with $\Y_{\pm}=(s_1, \dots, s_{r-1}, (\pm)^{d+1}\,s_r)$ an arbitrary $\so(d)$ integral
dominant weight and
\begin{itemize}
\item For $d=2r$: $\Delta=k-s_k$ for $k=1, \dots, r$ with $k=r$ only
  if $s_r \neq 0$;
\item For $d=2r+1$: $\Delta = \frac{d-2\ell}2$ with $\ell=1, \dots,
  r$ for bosonic irreps, or $\Delta = \frac{d+1-2\ell}2$ for
  fermionic irreps (with the same condition on $\ell$ as in the
  bosonic case).
\end{itemize}
The character of the module \eqref{nonintegral_module} is given by
\begin{equation}
  \chi^{\so(2,d)}_{\D(\Delta\,;\, \Y)}(q, \bm x) = \Big(q^\Delta\,
  \chi^{\so(d)}_{\Y_+}(\bm x) - q^{d-\Delta}\,
  \chi^{\so(d)}_{\Y_-}(\bm x) \Big)\, \Pd d (q, \bm x)\,.
  \label{sp FT ch}
\end{equation}
The scalar and the spinor order-$\ell$ singletons (respectively, Rac$_\ell$ and Di$_\ell$) are part of this
class of module \cite{Bekaert:2013zya}, as their conformal weight read
\begin{equation}
  \Delta = \frac{d-2\ell}2 +s\,,
\end{equation}
with $s=0$ or $s=\frac12$ (i.e. the spin of these fields). It is
clear from the above conditions that in odd dimension $d$, these two
families of singletons exhaust the fields described by this class of
module. However, for $d$ even one can consider more general conformal
fields (in particular they can be of spin greater or equal to one). In
general, the above class can be realized as conformal fields $\phi_\Y$
of spin $\Y$ obeying to a wave equation of the form:
\begin{equation}
  \partial^{d-2\Delta} \phi_\Y \approx 0\,.
\end{equation}
Notice that even if $\Y \neq \bm 0$, these fields do not enjoy any
gauge symmetry (contrarily to, for instance, the usual CHS fields
sourcing the conserved current in \eqref{noether_int}).

\paragraph{Character relations.}
Now that we have written down the character of the 
modules involved in the description of CHS fields, we can derive the
relation \eqref{rel_KT_FT_PM} used throughout the paper. Combining
\eqref{character_PM}, \eqref{character_Killing} and
\eqref{character_Weyl}, one can see that the three character, for
$d=2r$, are related by
\begin{eqnarray}
  \chi_{\D(1-s_1;(s_2-1,\dots, s_{p}-1, s_{p}-t, s_{p+1}, \dots,
      -s_r))}^{\so(2,d)}(q, \bm x) \qquad \qquad \qquad \qquad \qquad
  \qquad \qquad \label{KT_FT_PM} \\ \qquad \quad =\, (-1)^{p+1}\,
  \Big( \chi^{\so(2,d)}_{\D(d-\Delta_c;\Y_c)}(q, \bm x) +
  \chi^{\so(2,d)}_{\D(\Delta_{p}^{(t)};\Y_+)}(q, \bm x) -
  \chi^{\so(2,d)}_{\D(\Delta_{p}^{(t)};\Y_-)}(q^{-1}, \bm
  x)\Big)\,. \nonumber
  \label{gen rel mix}
\end{eqnarray}
On top of that, one can derive the following relation between
\eqref{character_PM}, \eqref{character_Weyl} and \eqref{character_Sh}
\begin{equation}
  \chi_{\Sh(d-\Delta^{(t)}_{p}; \Y)}^{\so(2,d)}(q, \bm x) =
  \chi^{\so(2,d)}_{\D(d-\Delta_c; \Y_c)}(q, \bm x) +
  \chi^{\so(2,d)}_{\D(\Delta^{(t)}_{p}; \Y)}(q, \bm x)\, ,
  \label{Sh_FT_PM}
\end{equation}
which, upon using the previously derived \eqref{KT_FT_PM}, also leads
to
\begin{equation}
  \chi_{\Sh(d-\Delta^{(t)}_{p}; \Y)}^{\so(2,d)}(q, \bm x) =
  (-1)^{p+1}\, \chi_{\D(1-s_1;(s_2-1,\dots, s_{p}-1, s_{p}-t,
      s_{p+1}, \dots, -s_r))}^{\so(2,d)}(q, \bm x) +
  \chi^{\so(2,d)}_{\D(\Delta_{p}^{(t)};\Y_-)}(q^{-1}, \bm x)\, .
  \label{Sh_KT}
\end{equation}
For $d=2r+1$ however, the relation reads
\begin{equation}
  \chi^{\so(2,d)}_{\D(1-s_1;(s_2-1,\dots, s_{p}-1, s_{p}-t,
      s_{p+1}, \dots, s_r))}(q, \bm x) = (-)^{p+1}\, \Big(
  \chi^{\so(2,d)}_{\D(d-\Delta_c; \Y_c)}(q, \bm x) +
  \chi^{\so(2,d)}_{\D(\Delta_{p}^{(t)}; \Y)}(q^{-1}, \bm x) \Big)\,.
\end{equation}
Recalling that for odd $d$ the character of the Weyl module is
identical to that of the shadow field, we can write the following
relation valid in any dimension:
\begin{eqnarray}
  && \chi^{\so(2,d)}_{\D(1-s_1;(s_2-1,\dots, s_{p}-1, s_{p}-t,
      s_{p+1}, \dots, (-1)^{d+1}s_r))}(q, \bm x) \\ && \qquad \qquad
  \qquad \qquad =\, (-1)^{p+1}\, \Big(
  \Sh^{\so(2,d)}_{\D(\Delta_{p}^{(t)};\Y)}(q, \bm x) +\, (-1)^{d+1}\,
  \chi^{\so(2,d)}_{\D(\Delta_{p}^{(t)}; \Y)}(q^{-1}, \bm x)
  \Big)\,. \nonumber
\end{eqnarray}
The above relation allows to express the character of the shadow field in terms of the character
of the associated partially massless field, plus or minus the
character of their conformal Killing tensor (depending on the symmetry
of these fields).

\paragraph{Action principles in the metric-like formulation.}
Let us conclude this appendix by commenting on the action principle
for a free conformal field of spin $\Y$ and conformal weight
$d-\Delta_{p}^{(t)}$. Schematically, it takes the form
\begin{equation}
  S_{\rm CHS}[\phi_\Y] \ = \ \int_{M_d} \dd^dx \ \phi_\Y\, \proj_{\sst\rm
    TT}^{\Y} \big(\partial^{2\Delta_{p}^{(t)}-d} \phi_\Y\big) \ =
  \ (-1)^{\Delta_{p}^{(t)}-\Delta_c} \int_{M_d} \dd^dx \ C_{\Y_c}\,
  \partial^{2\Delta_c-d} C_{\Y_c}\,.
  \label{action_sch}
\end{equation}
Notice that the number of derivatives involved in this action can be
constrained by requiring the latter to be quadratic in the shadow
field with spin $\Y$, and that the integrand has conformal weight
$d$. A careful analysis of such action principles can be found in
\cite{Vasiliev:2009ck}, where in particular the form of the action
\eqref{action_sch} is derived by requiring it to be conformally
invariant (see also \cite{Metsaev:2016oic} for the case of totally
symmetric fields).

\section{Two-dimensional case}
\label{app:d=2}

The $d=2$ conformal algebra is a direct sum of two $d=1$ conformal
algebras: $\so(2,2)=\so(2,1)\oplus \so(2,1)$ corresponding to left and
right movers if we write the flat metric in
light-cone coordinates $x^\pm$ on M$_2$ as $\dd s^2\, =\, 2\, \dd
x^+\, \dd x^-$\,. Accordingly, the spin label can have
positive/negative real values corresponding to left/right chirality. The
$\so(2,2)$ Verma modules will be denoted as $\cV(\Delta; s)$ and are
related to the $\so(2,1)$ Verma module $\cV_w$ with lowest weight
$w$ as
\begin{equation}
  \cV(\Delta;s) = \cV_{\frac{\Delta+s}2} \otimes
  \cV_{\frac{\Delta-s}2}
  \label{factor}
\end{equation}
Introducing the complex variables,
\begin{equation}\label{zzbar}
  z=q\, x=e^{-\b+i\,\a}\,, \qquad \bar z=q\, x^{-1}=e^{-\b-i\,\a}\,,
\end{equation}
for the $\so(2,2)$ weights,
the character of $\cV(\Delta;s)$ 
is given by
\begin{equation}
  \chi^{\so(2,2)}_{{\cal V}(\Delta;s)}(q,x) =
  \chi^{\so(2,1)}_{\cV_{\frac{\Delta+s}2}}(z)\,
  \chi^{\so(2,1)}_{\cV_{\frac{\Delta-s}2}}(\bar z) =
  \frac{z^{\frac{\Delta+s}2}\, {\bar
      z}^{\frac{\Delta-s}2}}{(1-z)(1-\bar z)}\,.
\end{equation}
The chirality-invariant modules will be denoted with $0$ as subscript,
e.g. ${\cal V}(\Delta;s)_0:={\cal V}(\Delta;+s)\oplus{\cal
  V}(\Delta;-s)\,$.

\paragraph{Conserved currents.}
The modules of conserved currents remain well-defined for $d=2$. In
particular, if one applies the definition \eqref{cons cur} in each
sector separately, then one gets
\begin{eqnarray}
  \cD(s;+s) &:=& \frac{\cV(s;+s)}{\cV(s+1;s-1)}\, \cong\, \cV_s\otimes
  1\\ \cD(s;-s) &:=& \frac{\cV(s;-s)}{\cV(s+1;1-s)}\, \cong\,1\otimes
  \cV_s
\end{eqnarray}
for the two chiralities, where $1$ stands for the trivial
representation of $\so(2,1)$ and we made use of \eqref{factor} and
$\cV_0/\cV_1\cong 1\,$. Accordingly, their respective characters
are
\begin{eqnarray}
  \chi^{\so(2,2)}_{\cD(s;+s)}(q,x) & = & \frac{z^s}{1-z}\, =
  \,\chi^{\so(2,1)}_{\cV_{s}}(z)\,,\\ \chi^{\so(2,2)}_{\cD(s;-s)}(q,x)
  & = & \frac{{\bar z}^s}{1-\bar z}\,=\,\chi^{\so(2,1)}_{\cV_{s}}(\bar
  z)\,.
\end{eqnarray}
The above can be straightforwardly generalized to the
partially-conserved currents: (see e.g. Section 6 of
\cite{Gwak:2015jdo}).

For $s\geqslant 2$, one can identify the parity-invariant modules
$\cD(s;s)_0:=\cD(s;+s)\oplus\cD(s;-s)$ as describing the spin-$s$
traceless conserved currents.  Indeed, the tracelessness implies that
all components of the type $J_{+-\cdots}$ vanish, in which case the
conservation condition becomes either $\partial_-J_{+\cdots+}= 0$ or
$\partial_-J_{-\cdots-}= 0$. This leads to
$J_{\pm\cdots\pm}=\varphi^{(s)}_\pm(x^\pm)$, where $\varphi_\pm^{(w)}$
are densities of weight $w$ in one dimension described as the
$\so(2,1)$ Verma module $\cV_w\,$.  The spins $s<2$ require a more
careful discussion (see e.g. Section 4.2 of \cite{Basile:2018dzi}) so
we will just mention that $\di=\cD(\frac12;\frac12)_0$ describes the
conformal spinor while $\cD(1;\pm 1)$ describe a left/right chiral
boson $J_\pm=\varphi^{(1)}_\pm(x^\pm)$ which both solves the
Klein-Gordon equation $\partial_+\partial_-J_\pm= 0$ (so it is a
submodule of Rac) and the conservation law
$\partial_-J_++\partial_+J_-=0$ (so it is a submodule of the complete
module describing the spin-1 conserved current).

\paragraph{Shadow fields.}
The modules $\cD(2;(s,s))$ are ill-defined for $d=2$ and $s>1$, in
accordance with the fact that Weyl tensors are identically vanishing
in two dimensions. In such case, one might replace the formula
\eqref{off/on} for the character of the shadow field by the analytic
continuation of the identity \eqref{off/KT} for $d=2$. By formally
applying \eqref{off/KT} one finds the intriguing relations
\begin{eqnarray}
  \chi^{\so(2,2)}_{\Sh(2-s;+s)}(q,x) & = & \frac{\bar z^s}{\bar z-1}\,
  =\, -\chi^{\so(2,1)}_{\cV_{s}}(\bar z)\,,
  \\ \chi^{\so(2,2)}_{\Sh(2-s;-s)}(q,x) & = & \frac{{z}^s}{z-1}\, =\,
  -\chi^{\so(2,1)}_{\cV_{s}}(z)\,.
\end{eqnarray}
which one can summarize as
\begin{equation}\label{summaryd=2}
  \chi^{\so(2,2)}_{\Sh(2-s; \pm s)} = -\chi^{\so(2,2)}_{\cD(s; \mp
    s)}\,.
\end{equation}
In fact, the shadow fields are pure gauge in $d=2$ and the negative
sign corresponds to the residual gauge symmetries.\,\footnote{This
  formula is in perfect agreement with the standard treatment of $d=2$
  case (cf. the appendix of \cite{Tseytlin:2013fca} for a detailed
  discussion) and in particular reproduces the partition function
  given in \cite{Beccaria:2014jxa} by setting $x=1$.}  Consider a
shadow field $h_s$ of spin $s\geqslant 2$, one can first reach the
traceless gauge where all components of the type $h_{+-\cdots}$
vanish. The residual gauge symmetry reads $\delta_\xi
h_{\pm\cdots\pm}=\partial_\pm\xi_{\pm\cdots\pm}$ and allows to fix the
reach the trivial gauge $h_s=0$. Even after this complete
gauge-fixing, the parameters of the residual gauge symmetry are
functions of one variable:
$\xi_{\pm\cdots\pm}=\varphi^{(1-s)}_\mp(x^\mp)$\,. This is nothing but
the higher-spin version of the usual infinite-dimensional enhancement
of conformal symmetries in $d=2$.

Using \eqref{inversion} for $d=1$, one can obtain the relation
\begin{equation}
  \chi^{\so(2,1)}_{\cV_{1-s}}(q^{-1}) = -
  \chi^{\so(2,1)}_{\cV_{s}}(q)\,.
\end{equation}
The contragredient of the $\so(2,1)$ lowest-weight Verma module
$\cV_{\Delta}$ is the $\so(2,1)$ highest-weight Verma module with
highest weight $-\Delta$, which we will denote
$\overline{\cV}_{\Delta}$. Its character is
$\chi^{\so(2,1)}_{\overline{\cV}_\Delta}(q) =
\chi^{\so(2,1)}_{\cV_\Delta}(q^{-1})$\,.  Hence, it may be tempting to
reinterpret the character formulae for the spin-$s$ shadow fields in
terms of the one for the contragredient $\so(2,1)$ Verma module with
highest weight $s-1$. In this sense, one can write the formal
equality: $\Sh(s;+s)=1\otimes\overline{\cV}_{-(s-1)}$ and
$\Sh(s;-s)=\overline{\cV}_{-(s-1)}\otimes 1$.

The list of the relevant parity-invariant $\so(2,2)$-modules and their
field-theoretical interpretations are summarized in
\hyperref[list2]{Table \ref{list2}}.
\begin{table}[h]
  \resizebox{0.99\textwidth}{!}{
    \begin{tabular}{|c|c|c|c|}
      \hline Modules & AdS$_3$ & CFT$_2$ & Equivalent descriptions
      \\ \hline\hline
 $\cD(\frac12;\frac12)_0$
      & Di & Conformal spinor & $(\cV_{\frac12}\otimes 1)\oplus
      (1\otimes \cV_{\frac12})$ \\[5pt]\hline
			$\cD(1;1)_0$ & $U(1)^{\otimes
        2}$ Chern-Simons & Chiral bosons & $(\cV_1\otimes 1)\oplus
      (1\otimes \cV_1)$\\
			& with Dirichlet behavior & &\\\hline
			$\Sh(1;1)_0$ & $U(1)^{\otimes
        2}$ Chern-Simons & Shadows of & $(\overline{\cV}_0\otimes 1)\oplus
      (1\otimes \overline{\cV}_0)$\\
			& with Neumann behavior & chiral bosons &\\\hline
      $\cD(s;s)_0$ & Massless spin-$s$ field & Conserved spin-$s$
      current & $(\cV_{s}\otimes 1)\oplus (1\otimes \cV_s)$ \\
& with Dirichlet behavior & &
			\\\hline
      $\Sh(s;s)_0$ & Massless spin-$s$ field & Shadow spin-$s$
      field & $(\overline{\cV}_{1-s}\otimes 1)\oplus (1\otimes \overline{\cV}_{1-s})$\\
& with Neumann behavior & &
\\\hline
      $\cD(1-s;s-1)_0$ & Killing tensor & Conformal  &
      $(\cD_{s-1}\otimes 1)\oplus (1\otimes \cD_{s-1})$\\
			&&Killing tensor&\\\hline
    \end{tabular}
  }
  \caption{\label{list2}List of relevant $\so(2,2)$ modules and their
    field-theoretical interpretations}
\end{table}

\paragraph{Conformal higher-spin gravity.}

Conformal higher-spin gravity is highly degenerate in two dimensions
since its action identically vanishes for $s \geqslant 2$ 
in a topologically trivial background while the
sector of low spin ($s=0,1$) is non-local. Nevertheless, its partition
function can be defined (see \cite{Tseytlin:2013fca,
  Beccaria:2014jxa}). As far as characters are concerned, one can
write $d=2$ versions of \eqref{off-shell version}. They rely on the
$d=2$ Flato-Fronsdal theorems (cf. section 4.2 in
\cite{Basile:2018dzi}). For instance, in the simpler type-B case,
\begin{equation}
  \left(\chi^{\so(2,2)}_{\di}\right)^2 = 2\,
  \chi^{\so(2,2)}_{\cD(1,0)} + \sum_{s=1}^\infty
  \chi^{\so(2,2)}_{\cD(s,s)_0}\,.
\end{equation}
This implies the  relation
\begin{equation}
  \sum\limits_{s=1}^\infty \chi^{\so(2,2)}_{\Sh(2-s,s)_0} =
  \sum\limits_{s=1}^\infty \chi^{\so(2,2)}_{\cD(1-s,s-1)_0} +
  \sum\limits_{s=1}^\infty \chi^{\so(2,2)}_{\cD(s,s)_0}\,.
  \label{off d=2}
\end{equation}
Strictly speaking, the relation for the full spectrum of type-B
off-shell CHS theory would involve a term $\cD(1,0)_0=2\,\cD(1,0)$ on
each side of \eqref{off d=2} which we omitted for simplicity.  Notice
that \eqref{off d=2} can also be derived from the fact that the
finite-dimensional spin-$s$ $\so(2,1)$-module can be seen as the
following quotient: $\cD_{s-1} \cong
\overline{\cV}_{1-s}/\overline{\cV}_{s}$. Using \eqref{summaryd=2} and
\eqref{off d=2}, one deduces
\begin{equation}
  \sum\limits_{s=1}^\infty \chi^{\so(2,2)}_{\Sh(2-s,s)_0} = -
  \sum\limits_{s=1}^\infty \chi^{\so(2,2)}_{\cD(s,s)_0} = \frac12 \sum
  \limits_{s=1}^\infty \chi^{\so(2,2)}_{\cD(1-s,s-1)_0}\,.
\end{equation}
One may define formally the character of the parity-symmetric spin-$s$
on-shell FT module as the analytic continuation of the one for the
quotient in the right-hand-side of \eqref{def_FT_mod} for $d=2$,
i.e. the difference $\chi^{\so(2,2)}_{\Sh(2-s,s)_0} -
\chi^{\so(2,2)}_{\cD(s,s)_0}$. By summing over all spins, one can say
that \eqref{main_identity} also holds in $d=2$.
		
\section{Branching rule for the Fradkin-Tseytlin module}
\label{app:branching}
In this appendix, we derive the branching rule for the
$\so(2,d)$-module of the on-shell FT field in even dimension $d=2r$, i.e.  we
decompose this irreducible $\so(2,d)$-module into a direct sum of
$\so(2,d-1)$-modules. The latter can naturally be interpreted as
fields in AdS$_d$, thereby allowing us to recover the factorization
property of the kinetic operator of CHS fields in M$_d$ into a
product of kinetic operators of partially-massless and massive fields
in AdS$_d$ obtained in \cite{Joung:2012qy, Nutma:2014pua,
  Metsaev:2014iwa}.

\subsection*{C.1. Branching rule for generalized Verma modules}
In order to derive the branching rules for on-shell Fradkin-Tseytlin
modules, we first need the following lemma \cite{Artsukevich:2008vy,
  Kobayashi2015}:
\begin{lemma}\label{lemma}
  The $\so(2,d)$ generalized Verma module
  \begin{equation}
    \V\big( \Delta \,;\, (s_1, \dots, s_r)\big) := \U\big( \so(2,d)
    \big) \otimes_{\U\big( (\,\so(2) \oplus \so(d)\,) \inplus\, \R^d
      \big)} \Vp_{[\Delta\,;\, (s_1, \dots, s_r)]}
    \label{def_gen_Verma}
  \end{equation}
  where $\Vp_{[\Delta\,;\, (s_1, \dots, s_r)]}$ is an $\so(2) \oplus
  \so(d)$ module (on which $\R^d$ acts trivially) of highest weight
  $[\Delta\,;\, (s_1, \dots, s_r)]$ branches onto $\so(2,d-1)$
  according to the following branching rule
  \begin{itemize}
  \item For $d=2r$
    \begin{equation}
      \V\big( \Delta \,;\, (s_1, \dots, s_r)\big) \quad \branch \quad
      \bigoplus_{n=0}^\infty \bigoplus_{\sigma_1=s_2}^{s_1} \dots
      \bigoplus_{\sigma_{r-1}=s_r}^{s_{r-1}} \V\big( \Delta+n \,;\,
      (\sigma_1, \dots, \sigma_{r-1})\big)\,;
      \label{branching_conf_even}
    \end{equation}
  \item For $d=2r+1$
    \begin{equation}
      \V\big( \Delta \,;\, (s_1, \dots, s_r)\big) \quad \branch \quad
      \bigoplus_{n=0}^\infty \bigoplus_{\sigma_1=s_2}^{s_1} \dots
      \bigoplus_{\sigma_r=-s_r}^{s_r} \V\big( \Delta+n \,;\,
      (\sigma_1, \dots, \sigma_r)\big)\,.
      \label{branching_conf_odd}
    \end{equation}
  \end{itemize}
\end{lemma}

The idea of the proof (for more technical details, see
e.g. \cite{Artsukevich:2008vy, Kobayashi2015}) is simple enough, when
expressed in terms of the Poincar\'e-Birkhoff-Witt basis, so that we
recall it here, before giving another proof in terms of characters.

\paragraph{Poincar\'e-Birkhoff-Witt basis.}
The generalized Verma module $\V(\Delta; \Y)$ defined by
\eqref{def_gen_Verma} is constructed as follows: one starts from the
module $\Vp_{[\Delta\,;\, \Y]}$ which is an irreducible representation
of the homothety subalgebra $\mathfrak{p}_d := (\,\so(2) \oplus
\so(d)\,) \inplus \R^d$ in which the action of $\R^d$ is
trivial. Concretely, the conformal algebra admits a three-grading
decomposition so that (as a vector space),
\begin{equation}
  \so(2,d) = \mathfrak{g}_{-1} \oplus \mathfrak{g}_0 \oplus
  \mathfrak{g}_{+1}\,,
\end{equation}
with
\begin{equation}
  \mathfrak{g}_{-1} = {\rm span}\{K_a\} \cong \R^d\,, \qquad
  \mathfrak{g}_0 = {\rm span}\{M_{ab},D\}\,, \qquad \mathfrak{g}_{+1}
  = {\rm span}\{P_a\} \cong \R^d\,,
\end{equation}
where $K_a$ are the special conformal transformation generators, $P_a$
the translation generators, $M_{ab}$ the $\so(d)$ generators and $D$
the dilation generator (while $a,b=1,\dots,d$). The vector space
$\Vp_{[\Delta;\Y]}$ carries a representation of $\mathfrak{p}_d$ in
the sense that $\so(2)$ acts diagonally with $\Delta$ as eigenvalue,
$\so(d)$ acts through the representation $\Y=(s_1,\dots,s_r)$ while
$\R^d \cong \mathfrak{g}_{-1} \cong {\rm span}(K_a)$ acts
trivially. In other words, it defines a primary field, i.e. a field
with fixed conformal weight $\Delta$ and spin $\Y$ annihilated by
special conformal transformations at the origin. The generalized Verma
module $\V(\Delta; \Y)$ is then constructed by acting with the whole
universal enveloping algebra $\U\big(\so(2,d)\big)$ on the
finite-dimensional module $\Vp_{[\Delta;\Y]}$, modulo the action of
$\mathfrak{p}_d$. In practice, the action of (arbitrary powers of) the
generators of $\mathfrak{g}_{+1} \cong \R^d$ on $\Vp_{[\Delta;\Y]}$
defines new vectors (or states) in the generalized Verma module, so
that
\begin{equation}
  \V\big( \Delta\,;\, \Y\big) \cong \U(\mathfrak{g}_{+1}) \otimes
  \Vp_{[\Delta; \Y]}\,.
\end{equation}
The action of the conformal algebra on $\V(\Delta\,; \Y)$ is then
defined by the action of the subalgebra $\mathfrak{p}_d$ on
$\Vp_{[\Delta;\Y]}$ together with the Lie bracket between generators
of this subalgebra and of $\mathfrak{g}_{+1}$. In other words, the
representation of $\so(2,d)$ is induced from representation of the
subalgebra $\mathfrak{p}_d$. In more concrete terms, we can write a
basis of $\V(\Delta;\Y)$ as
\begin{equation}
  \ket{\Delta+n\,; \Y}_{a_1 \dots a_n} = P_{a_1} \dots P_{a_n}
  \ket{\Delta\,; \Y}\,,
\end{equation}
where $\ket{\Delta\,;\Y}$ stands for a generic basis element of
$\Vp_{[\Delta; \Y]}$. In order to obtain the branching rule of
$\V(\Delta; \Y)$, the first task is to branch the representation
$\Vp_{[\Delta;\Y]}$ of $\mathfrak{p}_d$ to a direct sum of
representation of its lower dimensional counterpart,
$\mathfrak{p}_{d-1}$. Due to the fact that the $\R^d$ part acts
trivially and the $\so(2)$ is present in both $\mathfrak{p}_d$ and
$\mathfrak{p}_{d-1}$, we only need to consider the branching of the
$\so(d)$ component. For both $\so(2r)$ and $\so(2r+1)$, the branching
rules are well known and read respectively
\begin{equation}\label{Br1}
  (s_1, \dots, s_r) \quad
  \overset{\so(2r+1)}{\underset{\so(2r)}{\downarrow}} \quad
  \bigoplus_{\sigma_1=s_2}^{s_1} \dots
  \bigoplus_{\sigma_{r-1}=s_{r-2}}^{s_{r-1}}
  \bigoplus_{\sigma_r=-s_r}^{s_r} (\sigma_1, \dots, \sigma_r)\,,
\end{equation}
and
\begin{equation}\label{Br2}
  (s_1, \dots, s_r) \quad
  \overset{\so(2r)}{\underset{\so(2r-1)}{\downarrow}} \quad
  \bigoplus_{\sigma_1=s_2}^{s_1} \dots
  \bigoplus_{\sigma_{r-2}=s_{r-1}}^{s_{r-2}}
  \bigoplus_{\sigma_{r-1}=s_r}^{s_{r-1}} (\sigma_1, \dots,
  \sigma_{r-1})\,.
\end{equation}
The next thing to do is to decompose the action of
$\U(\mathfrak{g}_{+1}) \cong \U(\R^d)$ on the direct sum of the
representation of $\mathfrak{p}_{d-1}$ into a direct sum of
$\so(2,d-1)$ generalized Verma modules. To do so, let us single out
one generator of $\mathfrak{g}_{+1}$, say $P_d$, so that the remaining
generators $P_i$ ($i=1,\cdots, d-1$) span $\R^{d-1}$.  The basis of
$\V(\Delta;\Y)$ now reads
\begin{equation}
  P_{i_1}\dots P_{i_p} (P_d)^n \ket{\Delta; \bm \sigma}\,, \qquad n,
  p\in\N\,,\quad {\bm \sigma \in {\cal \bm B} (\Y)}\,,
\end{equation}
where the indices $i_1, \dots, i_p$ run from 1 to $d-1$, while
$\ket{\Delta; \bm \sigma}$ stands for a generic basis element of the
finite-dimensional $\mathfrak{p}_{d-1}$-module $\Vp_{[\Delta; \Y]}$
labeled by the $\so(2)$-eigenvalue $\Delta$ and $\so(d-1)$-weight $\bm
\sigma$, and ${\cal \bm B}(\Y)$ designates the set of
$\so(d-1)$-weights $\bm \sigma$ in the above branching rule of the
$\so(d)$ irrep $\Y $.  Due to the fact that $P_d$ has weight $+1$
under the adjoint action of the dilation generator (as any generator
of $\mathfrak{g}_{+1}$), and commutes with any element of $\so(d-1)$,
we have
\begin{equation}
  (P_d)^n \ket{\Delta;\bm\sigma} \cong \ket{\Delta+n;\bm\sigma}\,,
\end{equation}
i.e. it defines an irrep of $\mathfrak{p}_{d-1}$ with $\so(2)$-weight
$\Delta+n$ and $\so(d-1)$-weight $\bm\sigma\,$. Then the branching
rules \eqref{branching_conf_odd}-\eqref{branching_conf_even} of the
conformal algebra result from the branching rules
\eqref{Br1}-\eqref{Br2} of the rotation subalgebra.
	
\paragraph{Characters.}
The branching rules
\eqref{branching_conf_odd}-\eqref{branching_conf_even} can also be
recovered at the level of characters as follows:
\begin{itemize}
\item For $d=2r+1$, we have
  \begin{equation}
    \Pd d (q, \bm x) = \frac{1}{1-q}\, \Pd{d-1}(q, \bm x)\,,
  \end{equation}
  and
  \begin{equation}
    \chi^{\so(d)}_{(s_1, \dots, s_r)}(\bm x) = \sum_{\sigma=s_2}^{s_1}
    \dots \sum_{\sigma_r=-s_r}^{s_r}\, \chi^{\so(d-1)}_{(\sigma_1,
      \dots, \sigma_r)}(\bm x)\,,
  \end{equation}
  so that we readily obtain
  \begin{equation}
    \chi^{\so(2,d)}_{\V(\Delta;(s_1, \dots, s_r))}(q, \bm x) =
    \sum_{n=0}^\infty\, \sum_{\sigma=s_2}^{s_1} \dots
    \sum_{\sigma_r=-s_r}^{s_r}\,
    \chi^{\so(2,d-1)}_{\V(\Delta+n;(\sigma_1, \dots, \sigma_r))}(q,
    \bm x)\,,
    \label{branch_char_conf_odd}
  \end{equation}
  in accordance with \eqref{branching_conf_odd}.
    
\item For $d=2r$, the rule \eqref{branching_conf_even} is less
  straightforward to recover at the level of character due to the fact
  that the rank $\so(2,d)$ is $r+1$ whereas the rank of $\so(2,d-1)$
  is $r$, and hence the characters of modules of the latter depend on
  one less variables than the characters of the former. This can be
  already observed for the branching of $\so(d)$ irreps
  \eqref{branching_so_even} which at the level of characters reads
  \begin{equation}
    \chi^{\so(d)}_{(s_1, \dots, s_r)}(\bm x) + \chi^{\so(d)}_{(s_1,
      \dots, -s_r)}(\bm x) = \sum_{k=1}^r\, {\cal A}_k^{(r)}(\bm x)\,
    \sum_{\sigma=s_2}^{s_1} \dots \sum_{\sigma_{r-1}=s_r}^{s_{r-1}}\,
    \chi^{\so(d-1)}_{(\sigma_1, \dots, \sigma_{r-1})}(\bm x_k)\,,
      \label{branching_so_even}
  \end{equation}
  where
  \begin{equation}
    \bm x_k = (x_1, \dots, x_{k-1}, x_k, \dots, x_r)\,,
  \end{equation}
  and
  \begin{equation}
    {\cal A}_k^{(r)}(\bm x) = (x_k^{s_r}+x_k^{-s_r})\, \frac{\bm
      \delta(x_1+x_1^{-1}, \dots, x_{k-1}+x_{k-1}^{-1}, 2,
      x_{k+1}+x_{k+1}^{-1}, \dots x_r+x_r^{-1})}{\bm
      \delta(x_1+x_1^{-1}, \dots, x_r+x_r^{-1})}\,,
  \end{equation}
  where $\bm \delta(X_1, \dots, X_r)$ is the $r \times r$ Vandermonde
  determinant. In order to simplify \eqref{branching_so_even}, one can
  set one of the variables $\bm x$ of the $\so(d)$ character to $1$,
  say $x_r$, so that the characters on both sides of this equations
  depend on the same number of variables. Due to the fact that
  \begin{equation}
    {\cal A}_k^{(r)}(\bm{\check x}) = 2\, \delta_{k,r}\,, \qquad
    \text{with} \qquad \bm{\check x} := (x_1, \dots, x_{r-1}, 1)\,,
  \end{equation}
  and
  \begin{equation}
    \chi^{\so(d)}_{(s_1, \dots, s_{r-1}, s_r)}(\bm{\check x}) =
    \chi^{\so(d)}_{(s_1, \dots, s_{r-1}, -s_r)}(\bm{\check x})\,,
  \end{equation}
  the branching rule \eqref{branching_so_even} simplifies to
  \begin{equation}
    \chi^{\so(d)}_{(s_1, \dots, s_{r-1}, \pm s_r)}(\bm{\check x}) =
    \sum_{\sigma=s_2}^{s_1} \dots \sum_{\sigma_{r-1}=s_r}^{s_{r-1}}\,
    \chi^{\so(d-1)}_{(\sigma_1, \dots, \sigma_{r-1})}(\bm x_r)\,.
  \end{equation}
  Similarly, we have
  \begin{equation}
    \Pd d (q, \bm{\check x}) = \frac{1}{1-q}\, \Pd{d-1} (q, \bm
    x_r)\,,
  \end{equation}
  and hence
  \begin{equation}
    \chi^{\so(2,d)}_{\V(\Delta;(s_1, \dots, s_r))}(q, \bm{\check x}) =
    \sum_{n=0}^\infty\, \sum_{\sigma=s_2}^{s_1} \dots
    \sum_{\sigma_r=s_r}^{s_{r-1}}\,
    \chi^{\so(2,d-1)}_{\V(\Delta+n;(\sigma_1, \dots,
      \sigma_{r-1}))}(q, \bm x_r)\,,
    \label{branch_char_conf_even}
  \end{equation}
  in accordance with \eqref{branching_conf_even}.
\end{itemize}
This concludes the alternative proof via the computation of
characters.

\subsection*{C.2. Branching rule for irreducible lowest-weight modules}
Having the above \hyperref[lemma]{Lemma \ref{lemma}} at hand, we can
then obtain the branching rule of an irreducible $\so(2,d)$-module
$\D(\Delta;\Y)$ defined in terms of quotients of generalized Verma
modules. Indeed, as we have seen previously (in
\hyperref[app:zoo]{Appendix \ref{app:zoo}}), the characters of such
quotient are given by an alternating sum of characters of generalized
modules $\V(\Delta';\Y')$ as defined in \eqref{def_gen_Verma}, while
the branching rule of such a module translates straightforwardly at
the level of its character. Hence, applying the identity
\eqref{branch_char_conf_even} and \eqref{branch_char_conf_odd} to each
of the terms appearing in the character of an irreducible module
$\D(\Delta;\Y)$, and reinterpreting the resulting expression as a sum
of characters of irreducible modules of $\so(2,d-1)$, we can deduce
the branching rule for $\D(\Delta;\Y)$.

\paragraph{Fradkin-Tseytlin modules.}
As emphasized in the rest of the paper, the module of
central interest in CHS gravity is that of the on-shell
(totally-symmetric) FT field introduced in
\eqref{def_FT_mod}. Before deriving its branching, let us spell out
its character in all dimensions (taking $d=2r$ in the rest of the
section):
\begin{itemize}
\item The character of the $\so(2,d)$-module of the \textit{on-shell}
  FT field in even $d$ dimensions reads
  \begin{eqnarray}
    \chi^{\so(2,d)}_{\D(2; (s, s))}(q, \bm x) & = & \sum_{m=0}^{r-2}
    (-1)^m \Big( q^{2+m}\, \chi^{\so(d)}_{(s, s, \1^m_+)}(\bm x) +
    q^{d-2-m}\, \chi^{\so(d)}_{(s, s, \1^m_-)}(\bm x) \Big)\, \Pd d
    (q, \bm x) \nonumber \\ && \qquad \quad - 2 \big( q^{s+d-2}\,
    \chi^{\so(d)}_{(s)}(\bm x) - q^{s+d-1}\, \chi^{\so(d)}_{(s-1)}(\bm
    x) \big)\, \Pd d (q, \bm x)\,;
  \end{eqnarray}
\item The character of the $\so(2,d-1)$-module of the
  \textit{off-shell} FT field in odd $d-1$ dimensions reads
  \begin{eqnarray}
    \chi^{\so(2,d-1)}_{\D(2; (s, s-t+1))}(q, \bm x) & = &
    \sum_{m=0}^{r-3} (-1)^m \big( q^{2+m} - q^{d-3-m}\, \big)
    \chi^{\so(d-1)}_{(s, s-t+1, \1^m)}(\bm x)\, \Pd{d-1} (q, \bm x)
    \\ && \qquad + \big( q^{s+d-t-2}\, \chi^{\so(d-1)}_{(s)}(\bm x) -
    q^{s+d-2}\, \chi^{\so(d-1)}_{(s-t)}(\bm x) \big)\, \Pd{d-1}(q, \bm
    x)\,. \nonumber
  \end{eqnarray}
\end{itemize}
Now applying \hyperref[lemma]{Lemma \ref{lemma}} as explained in the
previous paragraph to the character of the Weyl module for $d=2r$
yields, after taking care of a few cancellations
\begin{eqnarray}
  \chi^{\so(2,d)}_{\D(2; (s, s))}(q, \bm{\check x}) & = &
  \sum_{t=1}^{s+\frac{d-4}2} \chi^{\so(2,d-1)}_{\D(s+d-t-2;(s))}(q,
  \bm x_r) + \chi_{\Sh(1+t-s;(s))}^{\so(2,d-1)}(q, \bm x_r)\,.
\end{eqnarray}
The characters appearing on the right-hand-side of the above equation
correspond to the following modules:
\begin{itemize}
\item For $t=1,\dots,s$, the characters
  $\chi^{\so(2,d-1)}_{\D(s+d-t-2;(s))}(q, \bm x_r)$ are those of the
  modules describing partially-conserved currents of spin-$s$ and
  depth-$t$ in $d-1$ dimensions (given in \eqref{character_PM_sym}),
  while $\chi_{\Sh(1+t-s;(s))}^{\so(2,d-1)}(q, \bm x_r)$ are the
  characters of the corresponding spin-$s$ and depth-$t$ shadow fields
  in M$_{d-1}$. Notice that the former/latter modules can also be
  interpreted as AdS$_d$ fields, namely as spin-$s$ and depth-$t$
  partially-massless fields with Dirichlet/Neumann boundary behavior.
\item For $t=s+1, \dots, s+\frac{d-4}2$, the characters
  $\chi^{\so(2,d-1)}_{\D(s+d-t-2;(s))}(q, \bm x_r)$ are those of the
  modules describing massive fields on AdS$_d$ with minimal energy
  $\Delta_+=d-3, \dots, \frac d2$, while
  $\chi_{\Sh(1+t-s;(s))}^{\so(2,d-1)}(q, \bm x_r)$ are the characters
  of spin-$s$ massive fields on AdS$_d$ with minimal energy
  $\Delta_-=2, \dots, \frac{d}2-1$. These modules come by pair of
  conjugate dimensions $\Delta_+$ and $\Delta_-=d-\Delta_+$
  corresponding to massive field with identical mass but Dirichlet vs
  Neumann boundary behavior.
\end{itemize}
Having in mind that the $\so(2,d-1)$-modules $\Sh(s+d-t-2;(s))$ and
$\D\big( 2; (s, s-t+1)\big)$ coincide for $t=1,\dots,s$ and correspond
to the off-shell FT module either in terms of the shadow field $h_s$
or in terms of its Weyl tensor $C_{s,s}$, we can write the branching
rule:
\begin{eqnarray}
  \D\big( 2; (s, s)\big) & \,\, \branch \,\, & \bigoplus_{t=1}^s
  \D\big( 2; (s, s-t+1)\big) \oplus \D\big( s+d-t-2; (s)\big) \nn &&
  \qquad \qquad \qquad \oplus \bigoplus_{n=0}^{d-5} \D\big( d-3-n;
  (s)\big)\,.
  \label{final_decompo_branching}
\end{eqnarray}
This decomposition is in accordance with the results of
\cite{Joung:2012qy, Nutma:2014pua, Metsaev:2014iwa} (conjectured in
\cite{Tseytlin:2013jya}) where the decomposition of the CHS wave
operator in $\R^d$ into a product of wave operators in AdS$_d$ for the
partially massless fields of depths $t=1,\cdots,s$\,.  Firstly, the
left-hand-side describes the on-shell FT field on $d$-dimensional
conformally flat space M$_d$ as a module of the conformal algebra
$\mathfrak{so}(2,d)$. Secondly, the right-hand-side correspond to its
description in terms of partially massless fields on AdS$_d$ as
modules of the isometry algebra $\mathfrak{so}(2,d-1)$. Thirdly, each
partially massless field gives rise to two modules: one for each
possible boundary choice. Following similar observations in
\cite{Beccaria:2016tqy}, let us rephrase the heuristic argument as
follows: The conformal boundary of Euclidean\,\footnote{We preferred
  the Euclidean signature because the geometric picture is more
  intuitive, but it holds in Lorentzian signature as well.}
AdS$_{d+1}$ is the conformally-flat sphere S$^d$. In turn, Euclidean
AdS$_d$ is conformal to one hemisphere of S$^d$ with two possible
choices of boundary conditions at the equator S$^{d-1}$. Therefore,
defining a conformal field on S$^d$ in terms of the one on Euclidean
AdS$_d$ we need to sum over the two boundary condition choices.

Notice that the case $d=4$ is special, as no massive nor shadow fields
appears in the branching rule, so that
\begin{equation}
  \D\big( 2\,;\, (s, \pm s)\big) \quad \branchmod{4}{3} \quad
  \bigoplus_{t=1}^{s} \D\big(s+2-t\,;\, (s)\big)\,.
\end{equation}
In fact, the modules $\D\big( 2; (s, s-t+1)\big)$ are absent since
finite-dimensional irreducible $\so(3)$-modules labelled by two-row
Young diagrams vanish (and hence these modules do not exist for
$\so(2,3)$).

\paragraph{Higher-depth Fradkin-Tseytlin modules.}
More generally, for a depth-$t$ FT field, the branching rule for $d=4$
reads
\begin{equation}
  \D\big( 2\,;\, (s, \pm(s-t+1))\big) \quad \branchmod{4}{3} \quad
  \bigoplus_{\sigma=s-t+1}^s \bigoplus_{\tau=\sigma-s+t}^{\sigma}
  \D\big(\sigma+2-\tau\,;\, (\sigma)\big)\,,
\end{equation}
i.e. partially massless fields of different spins (as well as
different depth) appear. Notice that this is in accordance with the
factorization of the partition function of maximal depth FT fields
derived in \cite{Beccaria:2015vaa}. In higher dimensions ($d=2r
\geqslant 6$) the branching rule also involves more modules (as in the
$t=1$ case), namely
\begin{eqnarray}
  & \D\big( 2;(s, s-t+1)\big) & \nn & \,\, \branch \,\, &
  \bigoplus_{\sigma=s-t+1}^s\, \bigoplus_{\tau=\sigma-s+t}^\sigma
  \D\big( 2; (\sigma, \sigma-\tau+1)\big) \oplus \D\big(
  \sigma+d-\tau-2; (\sigma)\big) \nn && \qquad \qquad \oplus
  \bigoplus_{\sigma=s-t+1}^s\, \bigoplus_{n=0}^{d-5} \D\big( d-3-n;
  (\sigma)\big) \\ & \cong & \bigoplus_{\sigma=s-t+1}^s\,
  \bigoplus_{\tau=\sigma-s+t}^{\sigma+\frac{d-4}2}
  \Sh\big(1+\tau-\sigma; (\sigma)\big) \oplus \D\big(\sigma+d-\tau-2;
  (\sigma)\big)\nonumber
\end{eqnarray}
Notice that this branching rule contains $2\,t\,(s-t+\frac{d-2}2)$
modules whereas the kinetic operator for the FT field contains
$s-t+\frac{d-2}2$ factors (see e.g. \cite{Grigoriev:2018mkp}).  The
kernel of each factor operator, labelled by $k$ ranging from $0$ to
$s-t+\frac{d-4}2$, corresponds to the module $
\bigoplus_{\sigma=s-t+1}^s\, \Sh\big(1+t+k-s; (\s)\big) \oplus
\D\big(s+d-t-k-2; (\s)\big)$\,.  For instance, the $d=4$ maximal-depth
($s=t$) FT fields have two-derivative kinetic operators but their
spectrum is made of maximal-depth PM fields of spin 1 to $s$ (see
Section of 3.4 of \cite{Barnich:2015tma} for the concrete example of
$s=t=2$ case).

\section{A primer on nonlinear CHS gravity}
\label{app:Segal}

The nonlinear action of type-A CHS gravity is defined as the logarithmically divergent part of the
effective action of a conformal scalar field in the background of all shadow fields \cite{Tseytlin:2002gz, Segal:2002gd}. As such, it remains a somewhat formal definition which becomes concrete only once it is computed perturbatively in the weak
field expansion around the conformally-flat vacuum solution. 

\paragraph{Formal operator approach.}
The tower of shadow fields $h_s(x)$ is conveniently packed in a generating function over phase space: $h(x,p)=\sum_s h^{\mu_1\cdots\mu_s}(x)\,p_{\mu_1}\cdots p_{\mu_s}$\,. Using Weyl calculus, the latter can be interpreted as the symbol of a Hermitian differential operator $\hat{H}(\hat{x},\hat{p})=\sum_s h^{\mu_1\cdots\mu_s}(x)\,\hat{p}_{\mu_1}\cdots \hat{p}_{\mu_s}+\,\cdots$ and the CHS action is the Seeley-DeWitt coefficient $a_{\frac{d}2}[\hat{G}]$ of the operator $\hat{G}=\hat{p}^2+\hat{H}$. The weak field expansion of the latter coefficient can be computed via standard techniques in quantum mechanics \cite{Segal:2002gd,Bekaert:2010ky}.
Several qualitative features of nonlinear CHS gravity are more easily understood within what Segal called the ``formal operator approach'' (which will be shortened here to ``formal approach'' for brevity) which consists in writing all
formulae in terms of operators rather than their symbols and treating these operators as large $n\times n$ matrices. 
This formalism is very useful for elucidating the structure of the theory but its field-theoretical interpretation is sometimes fragile. Only the perturbative formulation with symbols admitting power series expansion in momenta admits a clear interpretation as a local field theory. From the point of view of the effective action, the formal approach amounts to treat the conformal scalar field $\phi$ as if it were a large $n$-vector (for instance an element of the space ${\mathbb C}^\infty$ of infinite sequences with finitely-many non-vanishing complex entries\footnote{In mathematical terms, the vector space ${\mathbb C}^\infty$ is the direct limit of the  $\mathbb N$-filtration of vector spaces:
$$
0\times {\mathbb C}\hookrightarrow {\mathbb C}^2 \hookrightarrow \ldots \hookrightarrow {\mathbb C}^N \hookrightarrow {\mathbb C}^{N+1} \hookrightarrow \ldots
$$
The collection of the sequences with one entry equal to $1$, and all other entries vanishing, provides a countably-infinite basis of the vector space ${\mathbb C}^\infty$\,.}). Therefore, the formal approach is essentially\footnote{More precisely, the $U(N)$-vector model should be replaced with the large-$n$ limit of the $U(N)$-singlet sector of a $U(nN)$-vector model (see e.g. \cite{Bekaert:2012ux}).} equivalent to ignoring functional and locality issues, as well as polynomial UV-divergencies, in the field-theoretical discussions of Section \ref{sec:typeA}. In other words, ``formal CHS gravity'' should correspond to the limiting case $d\to 0$ in a suitable dimensional regularization of $d$-dimensional CHS theory (and, similarly, of its related Neumann MHS theory around AdS$_{d+1}$).

\paragraph{Formal group of symmetries.}
Using Weyl calculus, the action \eqref{noether_int} which is quadratic in the conformal scalar field can be rewritten in a compact way as
\be\label{freeact}
S[\phi; \{h_s\}_{s\in\N}] = \langle\phi|\,\hat{G}\,
|\phi\rangle\,.
\ee
It is manifestly invariant under the gauge transformations
\ba
&&|\phi\rangle\to\hat{O}^{-1}\,|\phi\rangle\,,\label{transf1}\\
&&\hat{G}\,\to\,\hat{O}^\dagger\,\hat{G}\,\hat{O}\,,\label{transf2}
\ea
where $\hat{O}$ is an invertible operator and $\hat{O}^\dagger$ stands for its Hermitian conjugate. 
The spectrum of off-shell CHS gravity is spanned by Hermitian differential operators $\hat{G}$ which are gauge fields with symmetries \eqref{transf2} and they couple to the free scalar field $\phi$ as the kinetic operator in \eqref{freeact}. 
In the formal approach, the gauge symmetries \eqref{transf2} form a group of invertible operators. These finite gauge symmetries \eqref{transf2} are actually generated by two Hermitian differential operators $\hat{S}$ and $\hat{X}$ as follows:
\be
\hat{O}=e^{\hat{S}+i\hat{X}}\,.\label{exponential}
\ee
They correspond, respectively, to higher-spin Weyl transformations 
\be\label{hsweyl}
\hat{G}\to\hat{O}\,\hat{G}\,\hat{O}\,,\qquad \hat{O}^\dagger=\hat{O}
\ee
for invertible and Hermitian operators $\hat{O}=e^{\hat{S}}$, and to higher-spin diffeomorphisms
\be\label{hsdiffs}
\hat{G}\to\hat{O}^{-1}\,\hat{G}\,\hat{O}\,,\qquad \hat{O}^\dagger=\hat{O}^{-1}
\ee
for unitary operators $\hat{O}=e^{i\hat{X}}$\,.\footnote{Notice that the space ${\mathbb C}^\infty$ carries the fundamental representation of the infinite general linear group $GL(\infty,{\mathbb C})$ which one can describe as the group of infinite invertible matrices with finitely-many non-vanishing complex entries outside the diagonal. Its Lie algebra is the Lie algebra $\mathfrak{gl}(\infty,{\mathbb C})$ of infinite matrices with finitely-many non-vanishing complex entries. In particular, the higher-spin diffeomorphisms may correspond to the adjoint action of the real subgroup $U(\infty)\subset GL(\infty,{\mathbb C})$ on its Lie algebra $\mathfrak{u(\infty)}$ of infinite Hermitian matrices with finitely-many non-vanishing entries.}
On the one hand, it is not clear whether such invertible operators \eqref{exponential} form a group because the products of such operators may not be the exponential of some \textit{differential} operators (i.e. of \textit{finite} order in the derivatives). This subtlety somewhat precludes a mathematically rigorous definition of the group of higher-spin gauge symmetries.
On the other hand, the infinitesimal transformations are perfectly well-defined because differential operators form an associative algebra under composition product.

\paragraph{Algebra of gauge symmetries.}
The infinitesimal form of the transformations \eqref{transf1}-\eqref{transf2} is
\ba
\delta|\phi\rangle&=&-(\hat{S}+i\hat{X})\,|\phi\rangle\,,\\
\delta\,\hat{G}&=&[\hat{G},\hat{S}]_++i\,[\hat{G},\hat{X}]_-\,,\label{fullinf} 
\ea
where $[\hat{A},\hat{B}]_\pm=\hat{A}\,\hat{B}\pm \hat{B}\,\hat{A}$ denotes the (anti)commutator.
The linearization of \eqref{fullinf} is 
\be\label{gequiv}
\delta\hat{H}=[\hat{p}^2,\hat{S}]_++i\,[\hat{p}^2,\hat{X}]_-
\ee
Let $\sigma(x,p)$ and $\xi(x,p)$ be the Weyl symbols of $\hat{S}(\hat{x},\hat{p})$ and $\hat{X}(\hat{x},\hat{p})$ with tensors $\sigma_{s-2}(x)$ and $\xi_{s-1}(x)$ as coefficients in the power series expansion in momenta. Then, the infinitesimal gauge transformation \eqref{gequiv} reproduces the form of the gauge equivalence \eqref{gauge tr} of shadow fields (up to signs and factors which we omit for simplicity).
In other words, the spectrum of \textit{linearized} off-shell CHS gravity (i.e. the tower of all shadow fields) can be encoded in a single Hermitian operator $\hat{H}$ modulo the gauge equivalence \eqref{gequiv}. 
In fact, the spectrum of \textit{nonlinear} off-shell CHS gravity can be defined as the space of Hermitian differential operators $\hat{G}$ which are (i) in the vicinity of the vacuum $\hat{G}_0=\hat{p}^2$ and (ii) given modulo the gauge symmetries \eqref{fullinf}.

\paragraph{Algebra of global symmetries.}
By definition, the global symmetries of CHS gravity preserve the vacuum, i.e. they correspond to operators $\hat{O}_0$ such that
\be
\hat{O}_0^\dagger\,\hat{G}_0\,\hat{O}_0=\hat{G}_0\,.
\ee 
The generators of such operators span the (C)HS algebra.
The finite transformations generated by such operators $\hat{O}_0$ are linear transformations of the perturbation
\be
\hat{H}\to\hat{O}_0^\dagger\,\hat{H}\,\hat{O}_0
\ee
since they preserve the vacuum by definition.
Its infinitesimal version defines the action of (C)HS algebra on the spectrum of free off-shell CHS gravity.
In fact, one can check that the gauge equivalence relation is preserved by this action. Therefore, the
spectrum of free \textit{off-shell} CHS gravity is a module of the higher-spin algebra.\footnote{Another simple way to reach this conclusion is to consider the pairing $\int d^dx\, h_s j_s$ between shadow fields $h_s$ and conserved currents $j_s$. The tower of currents spans the twisted-adjoint module. Each shadow field is the dual module of the current with respect to this pairing, therefore the tower of shadow fields is the dual of the twisted-adjoint module.} As a corollary, the spectrum of free \textit{on-shell} CHS gravity is also a module of the higher-spin algebra (because the tower of Bach tensors span a submodule isomorphic to the twisted-adjoint module, thus the quotient is also a module).

\paragraph{Formal action.}
In the formal approach, the action of CHS gravity is obtained by looking for a functional $W_{\sst\rm CHS}[\hat{G}]$ which must be invariant under the gauge symmetries \eqref{transf2}. The invariance under higher-spin diffeomorphisms \eqref{hsdiffs}, i.e. the adjoint action of unitary operators, implies that $W_{\sst\rm CHS}[\hat{G}]=$Tr$[F(\hat{G})]$ for some function $F$\,, while the   
invariance under higher-spin Weyl transformations \eqref{hsweyl} was argued by Segal to imply that $F(x)= C\,\Theta_0(x)+c\,\Delta(x)$
where $\Theta_0$ is the left-continous Heaviside step function (i.e. it vanishes at the origin) while $\Delta(x)$ is the discontinuous function vanishing for all $x$ except at the origin $x=0$ where it is equal to the unity.
The heuristic argument is as follows: since $\hat{G}$ is Hermitian, it can be diagonalized by a unitary transformation,
\be\label{unitransf}
\hat{U}^{-1}\,\hat{G}\,\hat{U}=\sum\limits_i \lambda_i\,\mid i\,\rangle\langle\, i\mid
\ee
where $\hat{U}^\dagger=\hat{U}^{-1}$ and the eigenvalues $\lambda_i$ are real numbers. Therefore,
Tr$[F(\hat{G})]=\sum\limits_i F(\lambda_i)$ for a function $F$ to be determined. Transformations \eqref{hsweyl} can rescale the eigenvalues by positive factors, hence
an invariant function $F(\lambda)$ can only depend on $\lambda$ through its sign. This essentially leaves the Heaviside step function as only possibility up to a normalization factor $C$ and a constant $c$ related to the value that one assigns at the origin
to the Heaviside function. Segal considered the choice $C=1$ and $c=\tfrac12$, i.e.
\be\label{Segalact}
S^{\sst\rm Segal}_{\sst \rm CHS}[\hat{G}]:=\mbox{Tr}[\Theta_{\frac12}(\hat{G})]\,.
\ee
where $\Theta_{\frac12}:=\Theta_0+\tfrac12\Delta=\tfrac12+\tfrac12$sgn (where ``sgn'' stands for the sign function).

\paragraph{Heat kernel expansion.}

A standard field-theoretical regularization of the one-loop effective action is the heat kernel prescription:
\be\label{heatdact}
W^{\sst\rm Heat}_{\Lambda}[\hat{G}]\,:=\,-\int^\infty_{\frac1{\Lambda}}\frac{dt}{t}\,\mbox{Tr}\Big[\exp\big(-t\,\hat{G}\big)\Big]\,,
\ee 
if one ignores IR divergencies.
For a semidefinite Hermitian differential operator of the form $\hat{G}=\hat{p}^2+\hat{H}$, the heat kernel expansion takes the form: 
\be
\mbox{Tr}[\exp(-t\,\hat{G})]=t^{-\frac{d}2}\sum\limits_{n=0}^\infty t^n\,a_n[\hat{G}] 
\ee
where $a_n$ are the Seeley-DeWitt coefficients. For $d$ even, one can say that the coefficient of the logarithmically divergent term identifies with the nonlinear CHS action:
\be\label{logindact}
S_{\sst\rm CHS}[\hat{G}]:= W^{\sst\rm Heat}_{\sst\log}[\hat{G}]=a_{\frac{d}2}[\hat{G}]=\frac1{2\pi i}\oint\frac{dz}{z}\, \mbox{Tr}[\exp(-z\,\hat{G})]\,.
\ee 
which can be seen as contour integral around the origin of the heat kernel.
If the trace of heat kernel $\mbox{Tr}[\exp(-z\,\hat{G})]$ 
is analytic in the region ${\rm Re}(z)> 0$ (except maybe for the neighborhood of the origin) and approaches to zero when $|z|\to \infty$ in this region, then
the contour integral in \eqref{logindact} can be modified as 
\be
	a_{\frac{d}2}[\hat{G}]=\frac1{2\pi i}\int\limits^{-i\,\infty-0^+}_{+i\,\infty-0^+}\frac{dz}{z}\, \mbox{Tr}[\exp(-z\,\hat{G})]\,.
	\label{D17}
\ee
The Heaviside step function admits an integral representation of analogous form,
\be
\Theta_{\frac12}(\lambda)=\frac{1}{2\pi\,i}\int\limits^{-i\,\infty-0^+}_{+i\,\infty-0^+}\frac{dz}{z}\,\exp(-z\lambda)\,.
\label{HS}
\ee
Therefore, it is tempting to identify $\mbox{Tr}[\Theta_\frac12(\hat{G})]$ with $a_{\frac{d}2}[\hat{G}]$\,, the logarithmically divergent piece of the effective action \eqref{D17},  in agreement with Segal's formal action \eqref{Segalact}.
However, one needs to be cautious here. For instance, if we boldly permute the contour integral with the trace in \eqref{logindact}, 
then we would be lead to consider
the identity function instead of the Heaviside:
\be
	1=\frac{1}{2\pi\,i}\oint\frac{dz}{z}\,\exp(-z\lambda)\,.
	\label{ID}
\ee
Then, we would identify the logarithmically divergent term  $a_{\frac{d}2}[\hat{G}]$ with $\mbox{Tr}(\hat{1})$\,. 
Indeed, even in the formal viewpoint taken by Segal, $\mbox{Tr}(\hat{1})$, namely the number $n$ of eigenvalues,
is invariant under higher-spin Weyl transformations \eqref{hsweyl}.
One may still think that 
$\mbox{Tr}(\hat{1})$ looks more weird than $\mbox{Tr}\big(\Theta(\hat G)\big)$.
But in fact both of them are very formal expressions as they give divergent series when summing over the eigenvalues.
The corresponding regularization is invisible in the integral representation \eqref{HS} and \eqref{ID} for individual eigenvalues.
Another way to regularize $\mbox{Tr}(\hat{1})$ is 
by regarding it as $\lim\limits_{z\to 0}\mbox{Tr}(\hat G^{-z})$,
which is known to give the logarithmically divergent term $a_{\frac{d}2}[\hat{G}]$ 
in the zeta function regularization method.
The reason for the formal equivalence between $\mbox{Tr}\big(\Theta(\hat G)\big)$ and $\mbox{Tr}(\hat{1})$ 
--- under the aforementioned conditions imposed on $\mbox{Tr}\big(\exp(-z\,\hat G)\big)$ ---
 is because the operator $\hat G$ under consideration is positive semidefinite
while the contribution of the zero modes is probably of zero measure.

\paragraph{Formal equation of motion.} 
Let us consider the equation of motion of CHS gravity derived from the action principle.
On the one hand, varying the operator $\hat G$  in the action \eqref{logindact} as
\be
	\delta S_{\rm\sst CHS}(\hat G)
	=-\frac1{2\pi i} \oint dz\,\mbox{Tr}(e^{-z\,\hat G}\,\delta\hat G)\,,
\ee
we obtain the equation of motion as
\be
	\frac1{2\pi i} \oint dz\,\la x_1|\,e^{-z\,\hat G}\,|x_2\ra\approx 0\,.
\ee
The above equation is bilocal depending on both $x_1$ and $x_2$, but we can reorganize them into the center position $x=\frac12(x_1+x_2)$
and the relative position $q=x_2-x_1$ taken as an auxiliary variable generating infinitely many higher-spin fields.
On the other hand, varying the operator $\hat G$  in the formal action \eqref{Segalact} would lead to the operator equation
\be\label{formaleqDirac}
\langle i_1|\,\delta(\hat{G})\,| i_2\rangle\approx 0\,.
\ee
Using \eqref{unitransf}, one finds the collection of equations $\delta(\lambda_i)\approx 0$ stating that the Hermitian operator $\hat{G}$ is non-singular in the sense that it has no normalizable eigenvector with vanishing eigenvalue, as was argued by Segal. This is indeed the case for  the vacuum solution $\hat{G}_0=\hat{p}^2$ in Euclidean signature since there are no normalizable solutions of Laplace equation. Therefore, one may also argue that the formal equation of motion is somewhat empty since all operators $\hat{G}=\hat{G}_0+\hat{H}$ in the close vicinity of the vacuum solution keep this property. This agrees with the alternative formal action $\mbox{Tr}(\hat{1})$. Their equivalence indeed relies on ignoring the contribution of zero modes.

\paragraph{Formal gauge fixing.} 
In the formal approach, one may then use the Cholevsky decomposition of positive-definite Hermitian matrices 
\be
\hat{G}\approx \hat{L}\hat{D}\hat{L}^\dagger\,,
\ee
where $\hat{D}$ can be any chosen diagonal matrix with positive entries and $\hat{L}$ is a lower-triangular invertible matrix with positive diagonal entries. Therefore, on-shell $\hat{G}$ can be mapped to the vacuum solution in the formal approach (setting $\hat{D}=\hat{G}_0$ and $\hat{L}^\dagger=\hat{O}$) by a (large) gauge transformation.

\bibliographystyle{jhep}
\bibliography{biblio}

\end{document}